\documentclass[a4paper,12pt]{article}

\usepackage{graphicx,amssymb,amsmath,epic,amsfonts,cite}

\newcommand{\be}{\begin{equation}}
\newcommand{\ee}{\end{equation}}
\newcommand{\bea}{\begin{eqnarray}}
\newcommand{\eea}{\end{eqnarray}}

\renewcommand{\Re}{\operatorname{Re}}
\renewcommand{\Im}{\operatorname{Im}}
\newcommand{\ol}[1]{\overline{#1}}

\newcommand\e{\mathrm{e}}
\newcommand\iu{\operatorname{i}}
\newcommand\diff{\mathrm{d}}

\begin{document}

\thispagestyle{empty}
\vspace*{.2cm}
\noindent
HD-THEP-08-1 \hfill 14 January 2008

\vspace*{1.5cm}
\begin{center}
{\Large\bf D7-Brane Motion from M-Theory Cycles\\[.4cm] 
and Obstructions in the Weak Coupling Limit}
\\[1.5cm]
{\large A.~P.~Braun$^a$, A.~Hebecker$^a$ and H.~Triendl$^b$}\\[.5cm]
{\it $^a$ Institut f\"ur Theoretische Physik, Universit\"at Heidelberg,
Philosophenweg 16 und 19\\ D-69120 Heidelberg, Germany}\\
{\it $^b$ II. Institut f{\"u}r Theoretische Physik der
  Universit{\"a}t Hamburg\\
Luruper Chaussee 149,  D-22761 Hamburg, Germany.}

{\small\tt (a.braun@thphys.uni-heidelberg.de,}
{\small\tt a.hebecker@thphys.uni-heidelberg.de,}\\
{\small and}\hspace*{.1cm} {\small\tt hagen.triendl@desy.de)}
\\[2cm]

{\bf Abstract}\end{center}
\noindent
Motivated by the desire to do proper model building with D7-branes and 
fluxes, we study the motion of D7-branes on a Calabi-Yau orientifold from 
the perspective of F-theory. We consider this approach promising since, 
by working effectively with an elliptically fibred M-theory compactification, 
the explicit positioning of D7-branes by (M-theory) fluxes is 
straightforward. The locations of D7-branes are encoded in the periods of 
certain M-theory cycles, which allows for a very explicit understanding of 
the moduli space of D7-brane motion. The picture of moving D7-branes on a 
fixed underlying space relies on negligible backreaction, which can be 
ensured in Sen's weak coupling limit. However, even in this limit we find 
certain `physics obstructions' which reduce the freedom of the D7-brane 
motion as compared to the motion of holomorphic submanifolds in the 
orientifold background. These obstructions originate in the intersections
of D7-branes and O7-planes, where the type IIB coupling cannot remain weak. 
We illustrate this effect for D7-brane models on $\mathbb{CP}^1\times
\mathbb{CP}^1$ (the Bianchi-Sagnotti-Gimon-Polchinski model) and on $\mathbb{CP}^2$. 
Furthermore, in the simple example of 16 D7-branes and 4 O7-planes on 
$\mathbb{CP}^1$ (F-theory on K3), we obtain a completely explicit 
parameterization of the moduli space in terms of periods of integral M-theory 
cycles. In the weak coupling limit, D7-brane motion factorizes from the 
geometric deformations of the base space. 

\newpage

\section{Introduction}

During the last years, significant progress has been made in the 
understanding of string-theoretic inflation, moduli stabilization, 
supersymmetry breaking and the fine tuning of the cosmological 
constant using the flux discretuum. The most studied and arguably best 
understood setting in this context is that of type IIB orientifolds 
with D3- and D7-branes~\cite{Giddings:2001yu} (which has close cousins in 
M-theory~\cite{Becker:1996gj, Gukov:1999ya,Dasgupta:1999ss}). 
Given this situation, it is 
clearly desirable to develop the tools for particle-phenomenology-oriented 
model building in this context. One obvious path leading in this direction 
is the study of the motion of D7-branes in the compact space and their 
stabilization by fluxes~\cite{Gorlich:2004qm,Lust:2005bd,Jockers:2004yj,
Jockers:2005zy,Gomis:2005wc}. The long-term goal must be to achieve sufficient 
control of D7-brane stabilization to allow for the engineering of the 
desired gauge groups and matter content based on the D7-brane open string 
sector~\cite{Watari:2004jg}. This is a non-trivial task since the underlying 
Calabi-Yau geometry has to be sufficiently complicated to allow for the 
necessary enormous fine tuning of the cosmological constant mentioned above. 

In the present paper we report a modest step towards this goal in the 
simple setting of the 8-dimensional Vafa model, where 16 D7-branes move 
on $T^2/Z_2$~\cite{Vafa:1996xn}. This 
motion can be viewed equivalently as the deformation of the complex structure 
of the dual F-theory compactification on K3. The relevant moduli space of 
D7-brane motion has recently been studied as part of the moduli space of 
K3$\times$K3 compactifications of F-theory to 4~dimensions (see, 
e.g.~\cite{Gorlich:2004qm,Lust:2005bd,Aspinwall:2005ad}).

One of our main results is the parameterization of the D7-brane motion on the 
compact space in terms of periods which are explicitly defined using the 
standard integral homology basis of K3. In other words, we explicitly 
understand the motion of the 16 D7-branes and of the background 
geometry in terms of shrinking or growing M-theory cycles stretched
between the branes or between the branes and the orientifold planes. 
In our opinion, this is a crucial preliminary step if one wishes to stabilize 
specific D7-brane configurations using fluxes (which, in this context, are 
inherited from M-theory fluxes and depend on the integral homology of K3). 
Further important points which we discuss in some detail in the following 
include the geometric implications of Sen's weak coupling limit~\cite{
Sen:1997gv}, the issue of obstructions arising when D7-brane motion is 
viewed from the type IIB (rather than F-theory) perspective, and the 
relevance of Sen's construction of the double-cover Calabi-Yau~\cite{
Sen:1997gv} to type IIB models with branes at singularities~\cite{
Cascales:2003wn,Verlinde:2005jr}.

Since the subsequent analysis is necessarily rather technical, we now
give a detailed discussion of the organization of the paper, stating
the main methods and results of each section. 

In Sect.~\ref{defang}, we begin with a discussion of the possible 
backreaction of D7-branes on the embedding space. This is of immediate concern 
to us since, in contrast to other D-branes, D7-branes have co-dimension 
2 and can therefore potentially modify the surrounding geometry 
significantly, even in the large volume 
limit~\cite{Greene:1989ya, Gibbons:1995vg, Bergshoeff:2006jj}. 
However, following the analysis of~\cite{Sen:1997gv}, it is possible to 
consider only configurations where Im\hspace{.5ex}$\tau\gg 1$ 
almost everywhere (weak coupling limit). It is easy to show that, in 
such a setting, the deficit angle of each brane always remains small 
($\sim 1/($Im$\hspace{.5ex}\tau)$), while each O-plane has deficit angle 
$\pi$. The solutions that contain a single D7-brane only develop a deficit 
angle at large distances away from the brane~\cite{Greene:1989ya,
Gibbons:1995vg, Bergshoeff:2006jj}. Sen's weak coupling limit assumes a 
compact space with proper charge cancellation 
between D-branes and O-planes, which allows for the possibility that no
deficit angle arises. One may say that, in contrast to other
models with branes on Calabi-Yau space, the D7-brane case is special in that
one has to take the weak coupling limit more seriously than the large volume
limit to be able to neglect backreaction. 

In Sect.~\ref{sec:obs} we discuss obstructions to D7-brane 
motion. For this purpose, we have to go beyond our simple model with base 
space $\mathbb{CP}^1$. Using the Weierstrass description of an 
elliptic fibration over $\mathbb{CP}^2$ or over $\mathbb{CP}^1\times 
\mathbb{CP}^1$ (which corresponds to the 
Bianchi-Sagnotti-Gimon-Polchinski model \cite{Bianchi:1989du,Gimon:1996rq}), it is 
easy to count the degrees of freedom of D7-brane motion. We find that the 
motion of D7-branes is strongly restricted as compared to the general motion 
of holomorphic submanifolds analysed in~\cite{Lerche:2002ck, Lerche:2002yw, 
Lerche:2003hs, Jockers:2004yj}. One intuitive 
way of understanding these `physics obstructions' is via the realization that 
D7-branes always have to intersect the O7-plane in pairs or to be tangent to 
it at the intersection point. We emphasize this issue since it serves as an 
important extra motivation for our approach via M-theory cycles: If the 
moduli space is described from the perspective of the M-theory complex 
structure, such obstructions are automatically included and no extra 
constraints on the possible motion of holomorphic submanifolds need to be 
imposed. 

Section~\ref{sec:K3} is devoted to a brief review of the geometry of K3.
This is central to our analysis as the moduli space of D7-branes on 
$T^2/Z_2$ (or, equivalently, the motion of 16 D7-branes and 4 O7-planes on 
$\mathbb{CP}^1$) is dual to the moduli space of M-theory on K3 in the limit 
where the K3 is elliptically fibred and the volume of the fibre torus is sent 
to zero. In this language, the weak coupling limit corresponds to sending the 
complex structure of the torus, which is equivalent to the type IIB 
axiodilaton $\tau$, to $i\infty$. We recall that the relevant geometric 
freedom is encoded entirely in the complex structure of K3, which is 
characterized by the motion of the plane spanned by Re\hspace{.5ex}$\Omega$ 
and Im\hspace{.5ex}$\Omega$ in a 20-dimensional subspace of 
$H_2(K3,\mathbb{R})$. Alternatively, the same information 
can be encoded in two homogeneous polynomials defining a Weierstrass model 
and thus an elliptic fibration. 

In Sect.~\ref{sgg}, we recall that at the positions of D7-branes the torus 
fibre degenerates while the total space remains non-singular. When two or 
more D7-branes coincide, a singularity of the total space develops, the 
analysis of which allows for a purely geometric characterization of the 
resulting ADE gauge symmetry. For the simple case of two merging branes, we
show explicitly that a homologically non-trivial cycle of K3 with the topology 
of a 2-sphere collapses~\cite{Sen:1996vd, Lerche:1999de}. This collapsing 
cycle is the basic building block 
which will allow us to parameterize the full moduli space of D7-brane motion
in terms of the periods of such cycles in the remainder of the paper.

In Sect.~\ref{gp}, we start developing the geometric picture of the 
D7-brane moduli space, which is one of our main objectives in the present 
paper. Our basic building block is the $S^2$ cycle stretched between two
D7-branes introduced in the previous section. Here, we construct this cycle 
from a somewhat different perspective: We draw a figure-8-shaped 1-cycle in 
the base encircling the two branes and supplement it, at every point, with 
a 1-cycle in the torus fibre. In this picture, it is easy to calculate the 
intersection numbers of such cycles connecting different D-brane pairs, 
taking into account also the presence of O-planes (see the figures in this 
section). The Dynkin diagrams of the gauge groups emerging when several 
branes coincide are directly visible in this geometric approach. In 
particular, the relative motion of four branes `belonging' to one of the 
O-plane can be fully described in terms of the above 2-brane cycles. The 
pattern of the corresponding 2-cycles translates directly in the Dynkin
diagram of SO(8). To obtain a global picture, we will have to supplement 
the cycles of these four SO(8) blocks by further cycles which are capable of 
describing the relative position of these blocks. 

Before doing so we recall, in Sect.~\ref{het}, the duality of F-theory on
K3 to the E$_8\times$E$_8$ heterotic string on $T^2$. This is necessary since 
we want to relate the geometrically constructed 2-cycles discussed above 
to the standard integral homology basis of K3, which is directly linked to 
the root lattice of E$_8\times$E$_8$. In particular, we explicitly identify 
the part of the holomorphic 2-form $\Omega$ which corresponds to the two Wilson
lines of the heterotic theory on $T^2$ and thus determines the gauge symmetry 
at a given point in moduli space. 

In Sect.~\ref{so8}, we start with the specific form of $\Omega$ which realizes 
the breaking of E$_8\times$E$_8$ to SO(8)$^4$ (corresponding to the choice of 
the two appropriate Wilson lines). The 2-cycles orthogonal to this particular 
$\Omega$-plane generate the root lattice of SO(8)$^4$ and can be identified 
explicitly with our previous geometrically constructed 2-cycles of the four 
SO(8) blocks. Thus, we are now able to express these 2-cycles in terms of the 
standard integral homology basis of K3. Geometrically, this situation 
corresponds to a base space with the shape of a pillowcase (i.e. $T^2/Z_2$)
with one O-plane and four D-branes at each corner. The remaining four 2-cycles 
of K3, which are not shrunk, can be visualized by drawing two independent 
1-cycles on this pillowcase and multiplying each of them with the two 
independent 1-cycles of the fibre torus. Thus, we are left with the task of 
identifying these geometrically defined cycles in terms of the standard 
homology basis of K3. The relevant space is defined as the orthogonal 
complement of the space of the SO(8)$^4$ cycles which we have already 
identified. We achieve our goal in two steps: First, we consider the smaller 
(3-dimensional) subspace orthogonal to all SO(16)$^2$ cycles (their shrinking 
corresponds to moving all D-branes onto two O-planes and leaving the two 
remaining O-planes `naked'). Second, we work out the intersection numbers with 
the $S^2$-shaped 2-cycles connecting D-branes from different SO(8) blocks. 
After taking these constraints into account, we are able to express the 
intuitive four cycles of the pillowcase in terms of the standard K3 homology 
basis.

Finally, in Sect.~\ref{dbp}, we harvest the results of our previous analysis 
by writing down a conveniently parameterized generic holomorphic 2-form 
$\Omega$ and interpreting its 18 independent periods explicitly as the 16 
D-brane positions, the shape of the pillowcase, and the shape of the 
fibre torus. Of course, the existence of such a parameterization of the 
moduli space of the type IIB superstring on $T^2/Z_2$ is fairly obvious and 
has been used, e.g., in~\cite{Gorlich:2004qm} and in the more detailed 
analysis of~\cite{Lust:2005bd}. Our new point is the explicit mapping between 
the periods and certain geometrically intuitive 2-cycles and, furthermore, 
the mapping between those 2-cycles and the standard integral homology basis 
of K3. We believe that this will be crucial for the future study of brane 
stabilization by fluxes (since those are quantized in terms of the 
corresponding integral cohomology) and for the generalization to 
higher-dimensional situations.

We end with a brief section describing our conclusions and perspectives 
on future work.

\section{Deficit angle of D7-branes}
\label{defang}

In this section we discuss the backreaction of D7-branes on the geometry. In 
general, D$p$-branes carry energy density (they correspond to black-hole 
solutions for the gravitational background~\cite{Johnson:2000ch}). For 
$p<7$, this deforms the geometry at finite distances, but the space remains 
asymptotically flat at infinity. Thus, backreaction can be avoided by 
considering D-brane compactifications in the large volume limit.

By contrast, objects with codimension two (such as cosmic strings or D7-branes)
produce a deficit angle proportional to their energy density. Thus, a D7-brane 
in 10 dimensions may in principle have a backreaction on the geometry which 
is felt at arbitrarily large distances. Let us first consider the effect of 
the energy density of a D7-brane. From the DBI-action
it is easy to see that the gravitational energy density of a D7-brane is 
proportional to $e^\phi = 1/\Im \tau \equiv 1/\tau_2$. A 
D7-brane is charged under the axion $C_0 = \Re \tau \equiv \tau_1$, and 
supersymmetry constrains $\tau$ to be a holomorphic function of the 
coordinates transversal to the D7-brane, see e.g.~\cite{Bergshoeff:2006jj}. 
As will become apparent in the following, this implies $\tau \to \iu \infty$ 
at the position of the D7-brane. Thus the energy density does not couple to 
gravity due to the vanishing of the string coupling near the D7-brane.

But there is another effect, first 
investigated in~\cite{Greene:1989ya, Gibbons:1995vg}, 
which is due to the coupling of D7-branes 
to the axiodilaton $\tau$. This coupling 
produces a non-trivial $\tau$ background 
around their position whose 
energy deforms the geometry. In flat 
space this effect produces a deficit angle
around the position of the D7-brane at 
large distance. Let us have a closer look at the case of 
finite distance and weak coupling, which 
is important for F-theory constructions in the weak coupling limit. 

The relevant part of the type IIB supergravity action is 
\begin{equation}
\label{action angle deficit}
\int \diff^{10} x \sqrt{g} \Big( R + 
\frac{\partial_\mu \tau\partial_\nu\bar{\tau}}
{2\tau^2_2}g^{\mu\nu} \Big) \ .
\end{equation}
The equations of motion (and supersymmetry) imply that $\tau$ is a 
holomorphic function of the coordinate $z$ parameterizing the plane 
transversal to the brane, $\tau=\tau(z)$. Because of the $SL(2,\mathbb{Z})$ 
symmetry of IIB string theory acting on $\tau$, it is helpful to use 
the modular function $j(\tau)$ instead of $\tau$ itself for the description 
of the dependence of $\tau$ on $z$. The function $j$ is a holomorphic 
bijection from the fundamental domain of $SL(2,\mathbb{Z})$ onto the Riemann 
sphere and is invariant under $SL(2,\mathbb{Z})$ transformations of $\tau$ 
(details can be found in~\cite{Chandrasekharan:1985}). 
Here we need the properties
\begin{align}
&j \sim \e^{-2 \pi \iu \tau} \qquad 
\textrm{for} \quad \tau \to \iu \infty \ ,
\label{j maps i infty}\\
&j( \e^{\,2\pi \iu/3}) = 0 \ ,
\label{j maps third root} \\
&j( \iu) = 1 \ .
\label{j maps i}
\end{align}
Since $\tau$ is holomorphic in $z$, the modular invariant 
function $j(\tau)$ depends holomorphically on $z$, and we can use the 
Laurent expansion of $j$ in $z$. As we encircle the D-brane at $z=0$, $j$ 
must encircle the origin once in the opposite direction 
(cf.~Eq.~(\ref{j maps i infty})). Thus $j$ must be proportional to $1/z$ and 
we can write
\begin{equation}
\label{j angle deficit}
j(\tau) \simeq \frac{\lambda}{z}
\end{equation}
for small $z$. Here $\lambda$ is a modulus, the overall scaling of the 
axiodilaton. 

From~\eqref{action angle deficit} we also deduce Einstein's equation
\begin{equation}
R_{\mu\nu}=\frac{1}{4\tau^2_2} \big(\partial_\mu\tau\partial_\nu\bar{\tau} + 
\partial_\nu\tau\partial_\mu\bar{\tau} \big) \ .
\end{equation}
Note that there is no term representing the energy density of the brane, 
as argued above. If we parameterize the plane orthogonal to the brane
by $z$ and write the metric as
\begin{equation}
ds^2 = \eta_{\mu \nu} \diff x^\mu \diff x^\nu + \rho(z,\bar{z})
\diff z \diff \bar{z} \ ,
\end{equation}
we finally arrive at the equation
\begin{equation}
\partial\bar{\partial}\ln \rho = \partial\bar{\partial}\ln \tau_2 \ ,
\label{diff eq angle deficit}
\end{equation} \par
which is solved by $\rho=\tau_2 f(z)\ol{f}(z)$. For the simplest case of 
a single D7-brane in infinite 10-dimensional space, one might expect that 
both $\tau_2$ and $\rho$ will not depend on the angle in the complex plane 
because of radial symmetry. However, this is not the case as we now explain. 

The simplest solution is given by declaring Eq.~(\ref{j angle deficit}) to 
be exact. Then $\tau(z)$ maps the transverse plane precisely once to the 
fundamental domain of $\tau$. Remember that the fundamental domain contains 
three singular points which are fixed points under some $SL(2,\mathbb{Z})$ 
transformation. These points are 

\begin{center}
\begin{tabular}[h]{c|c}
  $\tau$& invariant under\\  \hline
$\tau \to \iu \infty$&$T$\\
$\tau = \e^{\,2\pi \iu/3}$&$ST$\\
$\tau = \iu$ &$S$ \ ,
\end{tabular} 
\end{center}
where $S$ and $T$ are the standard generators of $SL(2,\mathbb{Z})$.
From \eqref{j maps i infty} and \eqref{j maps third root} we see 
that the first two of these points are mapped to $|j| \to \infty$ 
and $j = 0$. Thus by~\eqref{j angle deficit} these points are at 
the position of the brane and at infinity, respectively. But 
from~\eqref{j maps i} we see that the $S$-monodromy point is 
somewhere at finite distance and~\eqref{j angle deficit} 
tells us that this point sits 
at\footnote{This is of course 
only true approximately if we consider a
solution with many branes, so that~\eqref{j angle deficit} 
is the Laurent expansion around the position of a single brane.} 
$z = \lambda$. Thus the phase of $\lambda$ singles out a special 
direction and the radial symmetry is broken. 
Nevertheless, if $|\lambda|$ is very large compared to the 
region we are interested in, there is still an approximate radial symmetry, 
as can be seen from~\eqref{j maps i infty}.
The monodromy point at $z = \lambda$ does not 
deform the region near the brane (which is mapped
to large $\tau$) and the limit $|\lambda| \to \infty$ 
blows up the region where the radial 
symmetry is preserved. We will see later that this 
limit corresponds to the weak coupling limit of Sen~\cite{Sen:1997gv}. 

We now return to the generic case (where extra branes may be present and 
(\ref{j angle deficit}) is only approximate) and use the assumption that 
$|\lambda|$ is very large. As we approach a radially symmetric situation in 
this limit, we can neglect the angular derivatives 
in~\eqref{diff eq angle deficit} and arrive at
\begin{equation}
\frac{1}{r} \frac{\partial}{\partial r} \Big( r \frac{\partial \ln \rho}
{\partial r} 
\Big) =\frac{1}{r} \frac{\partial}{\partial r} \Big( r \frac{\partial \ln 
\tau_2}{\partial r} \Big)\,, \label{diff eq angle deficit 2}
\end{equation}
where $r$ is the radius in the $(z,\bar{z})$-plane.
The deficit angle is given by
\begin{equation}
\alpha = - \pi \cdot r \frac{\partial \ln \rho}{\partial r} \ .
\end{equation}
Inserting this into~\eqref{diff eq angle deficit 2}, we see that
\begin{equation}
\frac{\partial \alpha}{\partial r} = - \pi \cdot \frac{\partial}{\partial r} 
\Big( r \frac{\partial \ln \tau_2}{\partial r} \Big) \ .
\end{equation}
As discussed before, there is no energy density at the position of the brane 
and therefore the deficit angle is zero there. By integration we obtain
\begin{equation}
\alpha = - \pi \cdot r \frac{\partial \ln \tau_2}{\partial r} + \pi \cdot r 
\frac{\partial \ln \tau_2}{\partial r} \Big|_{r=0}\ .
\label{angle deficit alpha}
\end{equation}

Let us estimate the behavior near the brane, where $\tau \to \iu \infty$. 
From~\eqref{j maps i infty} we see that
\begin{equation}
\tau \simeq  \frac{\iu}{2 \pi} \ln j \simeq - \frac{\iu}{2 \pi} \ln \frac{z}
{\lambda} \ ,
\end{equation}
and therefore
\begin{equation}
\tau_2 \simeq - \frac{1}{2 \pi} \ln \big| \frac{z}{\lambda} \big| \ .
\label{angle deficit tau2 near brane}
\end{equation}
Thus~\eqref{angle deficit alpha} can be evaluated using
\begin{equation}
r \frac{\partial \ln \tau_2}{\partial r} \simeq 
\frac{\partial\ln(\ln(r/|\lambda|))}{\partial \ln r}=
\frac{1}{\ln(r/|\lambda|)}\simeq -\frac{1}{2\pi\tau_2}\,.
\end{equation}
Since $\tau_2\to\infty$ for $r\to 0$, the second term 
in~\eqref{angle deficit alpha} vanishes and we find
\begin{equation}
\alpha \simeq \frac{1}{2\tau_2} \ .
\end{equation}
We see that for $ r \ll | \lambda |$ this becomes small and therefore, in this 
limit, the deficit angle is small, too. Thus, we have derived quantitatively 
at which distances backreaction is small in the weak coupling limit. 

Away from the D7-brane the analysis presented above breaks down.
This is due to the monodromy point at $\lambda$
which destroys the radial symmetry of the configuration. To determine
the deficit angle that emerges at distances that are much larger than 
$|\lambda|$, one has to solve \eqref{diff eq angle deficit}. 
The solution, and thus the physics, depends on the
boundary conditions that are chosen. These are encoded in 
the shape of the function $f(z)$, which in turn is 
determined by the symmetry that is required of the
solution. The classic solution
of~\cite{Greene:1989ya, Gibbons:1995vg} which
argues for a deficit angle $\pi/6$, demands an $SL(2,\mathbb{Z})$ invariant 
and non-singular metric. The analysis of~\cite{Bergshoeff:2006jj} argues
that, due to its appearance in the definition of the Killing spinor, the 
function $f(z)$ should be invariant under the monodromy transformations of 
$\tau(z)$. This requirement introduces another $z$-dependent
factor which leads to an asymptotic deficit angle $2\pi/3$.

Our interpretation of this situation is as follows: In a configuration with 
a single D7-brane, one has only three monodromy points, $T$, $ST$ and $S$
in the complex plane. Without loss of generality, we can fix the 
$T$-monodromy point (i.e. the D7-brane) at zero and the $ST$-point at 
infinity. The definition of a deficit angle, which requires radial symmetry, 
is possible at distances from the brane much smaller or much larger than that 
of the $S$-monodromy point. In the first case the deficit angle is 
parametrically small, in the second case it is $2\pi/3$ 
following~\cite{Bergshoeff:2006jj}. Thus, there appears to be no room for
a deficit angle $\pi/6$.

\section{Sen's Weak Coupling Limit and its Consequences for D7-brane Motion}
\label{sec:obs}

F-theory is defined by a Weierstrass model on some K\"ahler manifold~\cite{ 
Vafa:1996xn}. The Weierstrass equation is
\begin{equation}
\label{weierstrass form}
y^2=x^3+fx+g \ ,
\end{equation}
where $f$ and $g$ are sections of the line bundles $L^{\otimes 4}$ and 
$L^{\otimes 6}$ respectively. The holomorphic line bundle $L$ is defined by 
the first Chern class of the base space:
\begin{equation}
c_1(L) = c_1(B) \ .
\label{chern class line bundle}
\end{equation}
This equation is derived from the Calabi-Yau condition of F-theory\footnote{
From
duality to M-theory we know that the compactification manifold of F-theory 
must be Calabi-Yau in order to preserve $\mathcal{N}=1$ supersymmetry.
}, 
cf.~\cite{Sen:1997gv}.

The brane positions are given by the zeros of the discriminant of the
Weierstrass equation \eqref{weierstrass form}
\begin{equation}
\label{Delta F-theory}
\Delta=4f^3+27g^2 \ .
\end{equation}
Before going to the weak coupling limit, these objects are $(p,q)$ branes 
which cannot be all interpreted as D7 branes simultaneously. Their 
backreaction on the geometry is strong~\cite{Vafa:1996xn}. Furthermore, 
the brane motion is constrained because the form of the homogeneous 
polynomial in (\ref{Delta F-theory}) is non-generic, i.e., the branes 
do not move independently.  

Let us discuss the weak coupling limit for F-theory compactifications, in 
which one can formulate everything in terms of D7-branes and O7-planes. 
Following~\cite{Sen:1997gv}, we parameterize
\begin{equation}
\label{def f}
f = C \eta - 3 h^2
\end{equation}
and
\begin{equation}
\label{def g}
g = h (C \eta - 2 h^2) + C^2 \chi \ ,
\end{equation}
where $C$ is a constant and $\eta$, $h$ and $\chi$ are homogeneous polynomials 
of appropriate degree (i.e. sections of $L^{\otimes n}$). Note that $f$ 
and $g$ are still in the most general form if we parameterize them as above. 
The weak coupling limit now corresponds to $C \to 0$. To see this, consider 
the modular function $j$ that describes the $\tau$ field\footnote{We have changed the
normalization of $j(\tau)$ in order to agree with the physics convention \cite{Sen:1997gv}. }:
\begin{equation}
j(\tau)=\frac{4 (24f)^3}{4f^3+27g^2}=\frac{4 (24)^3 (C\eta-3h^2)^3}{\Delta} \ .
\label{j weak coupling}
\end{equation}
The discriminant is given as
\begin{equation}
\Delta=C^2(-9h^2)(\eta^2+12h\chi)
\end{equation}
in the weak coupling limit. We observe that for $C \to 0$ we have 
$|j| \to \infty$ everywhere away from the zeros of $h$. Locally, this 
corresponds to the limit $\lambda \to \infty$ in~\eqref{j angle deficit}.
Furthermore, four pairs of branes merge to form the O-planes at 
the positions where $h=0$.

The remaining branes are the D7-branes of this orientifold model. Their 
position is defined by the equation
\begin{equation}
\eta^2 + 12 h \chi = 0 \ .
\label{brane obstructions}
\end{equation}
The left-hand side does not describe the most general section in the line 
bundle $L^{\otimes 8}$. We conclude that the D7-branes in an orientifold 
model do not move freely in general. These obstructions have, to our 
knowledge, so far not been investigated and are not included in the common 
description of orientifold models. We want to clarify this point further in 
the following.

As an explicit example we consider the Weierstrass model on $B \equiv 
\mathbb{CP}^1 \times \mathbb{CP}^1$. The reason why we take this example is 
that it has already been shown in~\cite{Sen:1997kw} that the degrees of 
freedom of the Weierstrass model are in complete agreement with the 
CFT-description of the T-dual orientifold model, the Bianchi-Sagnotti-Gimon-Polchinski 
model~\cite{Bianchi:1989du,Gimon:1996rq}. By counting the degrees of freedom 
we will show that this model has less degrees of freedom than a model of  
freely moving D-branes in the corresponding orientifold model. For this, we 
take the F-theory model described in~\cite{Sen:1997kw}, go to the weak 
coupling limit and recombine all O7-planes into a single smooth O7-plane 
wrapped on the smooth base space of the Weierstrass model. The double-cover 
Calabi-Yau space is now easily constructed. Furthermore, we recombine all 
D7-branes into a single smooth D7-brane eliminating all D7-brane 
intersections. We first assume, following~\cite{Jockers:2004yj} that this 
D7-brane moves freely as a holomorphic submanifold (respecting, of course, 
the $Z_2$ symmetry of the Calabi-Yau). This allows for a straightforward 
determination of the corresponding number of degrees of freedom from its 
homology. We will then compare this number with the degrees of freedom that 
are present in the actual F-theory model.

Let us investigate the F-theory model in greater detail. We call the 
homogeneous coordinates $[x_1:x_2]$ for the first $\mathbb{CP}^1$, and 
$[y_1:y_2]$ for the second $\mathbb{CP}^1$. By $x$ and $y$ we denote the 
generators of the second cohomology, where $x$ corresponds to the cycle that 
fills out the second $\mathbb{CP}^1$ and is pointlike in the first 
$\mathbb{CP}^1$, and vice versa for $y$. In a product of complex projective 
spaces a section in a holomorphic line bundle corresponds to a homogeneous 
polynomial. We want to determine the degree of the homogeneous polynomials 
that will be involved in the calculation. For this purpose, we calculate the 
first Chern class of the base space:
\begin{equation}
c_1(B) = 2x+2y \ .
\end{equation}
From~\eqref{chern class line bundle} we conclude that this is the first Chern 
class of the line bundle $L$. Thus, sections in $L$ correspond to homogeneous 
polynomials of degree (2,2) and $h$, being a section in $L^{\otimes 2}$, is 
a homogeneous polynomial of degree $(4,4)$. A generic polynomial of degree 
$(4,4)$ in two complex coordinates is irreducible. Thus, a generic $h$ 
indeed describes one single O7-plane that wraps both $\mathbb{CP}^1$s four 
times. Similarly, $\eta$ is a homogeneous polynomial of degree $(8,8)$, and 
$\chi$ is of degree $(12,12)$. The left-hand side of Equation~\eqref{brane 
obstructions} is then a homogeneous polynomial of degree $(16,16)$. This 
polynomial is generically also irreducible and describes a single 
D7-brane that wraps both $\mathbb{CP}^1$s sixteen times. Note that both $h$ 
and $\eta^2 + 12 h \chi$ do not have any base locus and thus define smooth 
hypersurfaces in $\mathbb{CP}^1 \times \mathbb{CP}^1$ by Bertini's theorem, 
cf.~\cite{Hubsch:1992nu}. 

Equation~\eqref{brane obstructions} defines an analytic hypersurface $S$ of 
complex dimension one, which is the position of the D7-brane. This is just a 
Riemann surface, and in order to identify its topology, it suffices 
to determine its Euler number. By the methods of~\cite{Hubsch:1992nu, 
Griffiths:1978} it is easy to calculate the Euler number of a hypersurface 
defined by a homogeneous polynomial. The Euler characteristic is given as
\begin{equation}
\chi(T(S)) = \int_S c_1(T(S)) \ ,
\end{equation}
The first Chern class of $T(S)$ is given in terms of the Chern classes
of the normal bundle of $S$ in $B$ and the tangent bundle of $B$ 
by the second adjunction formula:
\begin{equation}
c_1(T(S)) = c_1(T(B)) - c_1(N(S)) \ . 
\end{equation}	
The normal bundle of $S$ in $B$ is equivalent to
the line bundle that defines $S$ through one of
its sections\footnote{Note that this means that 
$c_1(N(S))$ defines a 2-form on $B$.}. 
Putting everything together, we arrive at:
\begin{align}
\chi(T(S)) &= \int_S c_1(T(S)) = \int_S c_1(T(B)) - c_1(N(S)) \nonumber \\ 
&=\int_B \left( c_1(T(B)) - c_1(N(S)) \right) \wedge c_1(N(S)) \ .
\end{align}
The Chern class of a line bundle on $\mathbb{CP}^1 \times \mathbb{CP}^1$
that has sections which are homogeneous polynomials of degree $(n,m)$
is simply $n x+m y$. Together with $c_1(T(\mathbb{CP}^1 \times 
\mathbb{CP}^1))=2x+2y$ we find that
\begin{equation}
\chi(T(S)) =\int_B  ((2-n)x+(2-m)y)\wedge (n x+ my) = 2(n+m-nm) \ .
\end{equation}
Here we used the relations $\int x\wedge y= 1$ and $\int x \wedge x = \int y 
\wedge y = 0$. By the simple relation $\chi(S)=2-2g$ we can now compute the 
genus of $S$, and therefore the Hodge number $h^{(1,0)}(S)$ to be\footnote{
One might be worried that setting $m$ or $n$ equal to zero, one can get a 
surface with $g<0$. However, a homogeneous polynomial of degree $(n,0)$ is 
reducible for $n >1$, corresponding to a collection of disconnected surfaces.
}
\begin{equation}
h^{(1,0)}(S) = g = (n-1)(m-1) \ .
\end{equation}

From~\cite{Jockers:2004yj} we know that the number of holomorphic 1-cycles 
of a D7-brane in the Calabi-Yau space that are odd under the orientifold 
action is equal to the number of its valid deformations\footnote{
Note 
that in~\cite{Jockers:2004yj} this was shown for the case of a Calabi-Yau 
3-fold compactification down to four dimensions. Here we consider 
compactifications on $K3$ down to 6 dimensions. Because the holomorphic 
$n$-form then has one leg less, the degree of the relevant cohomology is 
reduced by one as well.
}. 
Above we computed the number of holomorphic 1-cycles that are even under the 
orientifold action. To compute the number of holomorphic 1-cycles that are 
odd under the orientifold projection, we construct the double cover of the 
brane. The double cover of $B$ is a hypersurface in a 
$\mathbb{CP}^1$-fibration over the base $B$, which is defined 
by~\cite{Sen:1997gv}
\begin{equation}
\xi^2 = h \ ,
\end{equation}
where $\xi$ is the coordinate in one patch of $\mathbb{CP}^1$. This introduces 
branch points at the location of the O-planes. To get back to the orientifold, 
one then has to mod out the symmetry $\xi \to - \xi$. This again gives the 
orientifold with the topology of $B$ and with O7-planes at the zeros of $h$.

\begin{figure}[!ht]
\begin{center}
\includegraphics[height=3.5cm,]{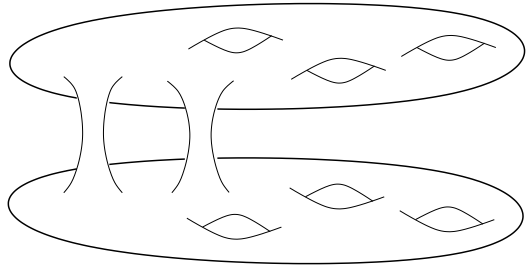}\caption{\textit{Illustration 
of the (double cover of) a D7-brane intersecting an O7-plane in four
points.}}
\label{dcover}
\end{center}
\end{figure}

We thus have to branch $S$ over the intersection points of the O7-plane with 
the D7-brane. The double cover $\tilde{S}$ of $S$ is then formed by two copies 
of $S$ that are joined by a number of tubes that is half of the number of 
intersections. We can compute the number of intersections between the D7-brane 
and the O7-plane by using the cup product between the corresponding elements 
in cohomology:
\begin{equation}
I_{(D7,O7)}=\int (4x+4y)\wedge (nx+my)=4(n+m) \ .
\label{obstructions intersections}
\end{equation}
To compute $h^{(1,0)}$ of $\tilde{S}$, we simply compute its genus (which is
equal to the number of handles of $\tilde{S}$). Because we are considering the
double cover, the genus of $\tilde{S}$ is twice the genus of $S$, plus a 
correction coming from intersections between the D7-brane and the O7-plane. 
As the intersections are pairwise connected by branch cuts which connect
$S$ to its image under the orientifold action, they introduce $\frac{1}{2} 
I_{(D7,O7)} - 1$ extra handles of $\tilde{S}$. This is illustrated in
Fig.~\ref{dcover}. Thus $h^{(1,0)}$ of $\tilde{S}$ is given by
\begin{equation}
h^{(1,0)}(\tilde{S})=g(\tilde{S})= 2g(S) + \frac{1}{2}I_{(D7,O7)} - 1\,.
\end{equation}
Now we can compute the number of holomorphic 1-cycles that are odd under the 
orientifold projection:
\begin{equation}
h_-^{(1,0)}(\tilde{S}) = h^{(1,0)}(\tilde{S}) - h^{(1,0)}(S) =  
g(S) + \frac{1}{2}I_{(D7,O7)} - 1 =(n+1)(m+1) - 1 \ .
\end{equation}
Thus a freely moving D7-brane in an orientifold model, defined by a 
homogeneous polynomial of degree $(n,m)$, has $h_-^{(1,0)}=(n+1)(m+1) - 1$ 
degrees of freedom. This fits nicely with the number of deformations of an 
arbitrary homogeneous polynomial of degree $(n,m)$. 

Coming back to our example, we see that a freely moving D7-brane on 
$\mathbb{CP}^1\times\mathbb{CP}^1$ corresponds to a homogeneous polynomial 
$P$ of degree $(16,16)$, giving $288$ degrees of freedom. 
From the F-theory perspective, this polynomial is constrained to be of the form
\begin{equation}
P = \eta^2 + 12 h \chi \ 
\label{obstructions brane equation}
\end{equation}
in the weak coupling limit. Here $h$ is fixed by the position of the O-plane. 
Let us count the degrees of freedom contained in this expression: $\eta$ is 
of degree $(8,8)$, and thus has $81$ degrees of freedom. $\chi$ is of degree 
$(12,12)$ and thus has $169$ degrees of freedom. Furthermore, we can always 
factor out one complex number from this equation so that we have to substract 
one degree of freedom. Finally, there is some redundancy that arises through 
polynomials $K = h^2 \alpha^2$ that are both of the form $\eta^2$ and $12 h 
\chi$. As $\alpha$ is a polynomial of degree $(4,4)$,
there are 25 redundant degrees of freedom in the present case. 
Putting everything together, we 
find that \eqref{obstructions brane equation} describes $224$ degrees of 
freedom. This differs by $64$ degrees of freedom from what we have found for 
an unconstrained polynomial of degree $(16,16)$. Note that this is just half 
of the number of intersections between D7-branes and O7-planes on 
$\mathbb{CP}^1\times\mathbb{CP}^1$.

Before discussing the local nature of these `physics obstructions' (which are 
very different from certain `mathematical obstructions' restricting the 
motion of holomorphic submanifolds in specific geometries~\cite{Kodaira1,
Brunner:1999jq,Lerche:2002yw,Jockers:2004yj}), we want to briefly review a 
second example. We take the base to be $\mathbb{CP}^2$, so that there are 
$6$ O7-planes and $24$ D7-branes in the weak coupling limit. We denote the 
only 2-cycle of $\mathbb{CP}^2$, which is a $\mathbb{CP}^1$, by $x$. After 
recombination we have one D7-brane that wraps the cycle $24x$ and one O7-plane
on the cycle $6x$. From $x \cdot x=1$ we conclude that they intersect
$144$ times. Now we can repeat counting the degrees of freedom contained in an
unconstrained polynomial of degree $24$, which is $324$, and compare it to
a polynomial of the form 
\eqref{obstructions brane equation}. For $\mathbb{CP}^2$, $h$ is of degree 
$6$, $\chi$ is of degree $18$ and $\eta$ is of degree $12$. By the same 
arguments as above we find that $252$ degrees of freedom are contained in 
\eqref{obstructions brane equation} for $\mathbb{CP}^2$. The two numbers
differ by $72$, which is again half of the number of intersections between
D7-branes and O7-planes.

It is then natural to expect that the obstructed D7-brane deformations are 
related to the intersections between D7-branes and O7-planes. Indeed, the 
smallness of the coupling cannot be maintained in the vicinity of O7-planes. 
Thus, our argument that the backreaction of D7-branes on the geometry is 
weak breaks down and we have no right to expect that D7-branes move freely 
as holomorphic submanifolds at the intersections with O7-planes. 

At the level of F-theory, the above physical type-IIB-argument is reflected in 
the non-generic form of the relevant polynomials in the weak coupling limit. 
To see this explicitly, let us investigate~\eqref{obstructions brane equation} 
in the vicinity of an intersection point. We parameterize the neighborhood of 
this point by complex coordinates $z$ and $w$. Without loss of generality, 
we take $h=w$ (i.e. the O7-plane is at $w=0$) and assume that the intersection 
is at $z=w=0$. This means that $P=(\eta^2 + 12 h \chi)$ vanishes at $z=w=0$ 
and, since we already know that $h$ vanishes at this point, we conclude that 
$\eta(z=0,w=0) = 0$. Expanding $\eta$ and $\chi$ around the intersection 
point, 
$$\eta(z,w) = m_1 z + m_2 w + \dots\qquad\mbox{and}\qquad\chi(z,w) = n_0 + 
n_1 z + n_2 w + \dots \,,$$
we find at leading order\footnote{
Note 
that $zw$ is subdominant w.r.t. to $w$, which is not true for $z^2$.
}
\begin{equation}
P = m_1^2 z^2 + 12n_0 w + \dots = 0 \ .\label{pexp}
\end{equation}
In the generic case $n_0 \ne 0$, this is the complex version of a parabola
`touching' the O-plane with its vertex. Thus, we are dealing with a double 
intersection point.\footnote{
This 
is also clear from the fact that, if we were to introduce by hand a term 
$\sim z$ in (\ref{pexp}), our intersection point would split into two.
} 
In the special case $n_0=0$, our leading-order $P$ is reducible and we are 
dealing with two D7-branes intersecting each other and the O7-plane at the 
same point. The former generic case hence results from the recombination of 
this D7-D7-brane intersection. In both cases, we have a double 
intersection point. In other words, the constraint corresponds to the 
requirement that all intersections between the D7-branes and O7-planes must 
be double intersection points. There are two easy ways to count the number 
of degrees of freedom removed by this constraint: On the one hand, demanding 
pairwise coincidence of the $2n$ intersection points (each of which 
would account for one complex degree of freedom for a freely moving 
holomorphic submanifold) removes $n$ degrees of freedom. On the 
other hand, at each of the $n$ double intersection points the coefficient of 
the term $\sim z$ in (\ref{pexp}) must vanish, which also removes $n$ 
complex degrees of freedom. 

To summarize, we have again arrived at the conclusion that $2n$ intersections 
between a D7-brane and an O7-plane remove $n$ of the degrees of freedom of 
the D7-brane motion. In particular, we have now shown that these `physics 
obstructions' have a local reason: In the weak coupling limit, the O7-plane 
allows only for double intersection points.\footnote{
This 
means that some of the 1-cycles that would be present in the double 
cover of a generic holomorphic submanifold are collapsed in the double cover 
of a D7-brane in the weak coupling limit. These are the 1-cycles that 
wind around two branch points.
}
At the local level, our findings are easily transferred to compactifications 
to 4 dimensions: The O7-plane D7-brane intersections are now complex curves 
rather than points. At each point of such a curve, we can consider the 
transverse compact space, which is again complex-2-dimensional. In this 
space, we can perform the same local analysis as above and conclude that
double intersection points are required. 

We end this section with a comment on an interesting application of 
F-theory and elliptic fibrations which may be useful for type-IIB model 
building on the basis of local Calabi-Yau constructions~\cite{
Cascales:2003wn,Verlinde:2005jr}: Consider a non-compact K\"ahler manifold 
with SU(3) holonomy (a local Calabi-Yau). Such spaces play an important 
role in attempts to construct Standard-like models from branes at 
singularities. We will now sketch a generic procedure allowing us
to embed them in a compact Calabi-Yau.

Assuming that the non-compact K\"ahler manifold is given as a toric variety, 
it is clearly always possible to make it compact by adding appropriate cones. 
Furthermore, this can always be done in such a way that the resulting 
compact K\"ahler manifold $B$ has a positive first Chern class.\footnote{
The 
fan of a toric Calabi-Yau is spanned by one-dimensional cones that
are generated by vectors ending on a single hyperplane $H$. If we add a
one-dimensional cone in the direction opposite to the normal vector $n_{H}$
of $H$,
we end up with a (in general) non-compact K\"ahler manifold of positive
first Chern class. To make this space compact, we appropriately enlarge the
fan. As this can always be done such that all the one-dimensional cones that
are added are generated by vectors that end on $H$, we do not have to change
the first Chern class.
}
With the 
first Chern class we can associate a line-bundle $L$ and a divisor. Positivity 
of the Chern class implies that this divisor is effective, i.e., the line 
bundle $L$ has sections without poles (the zero locus of such a section
defines the divisor). Now we can wrap an O7-plane (with four D7-branes on 
top of it) twice along the above effective divisor. This corresponds to the 
orientifold limit of a consistent F-theory model. Indeed, as explained at 
the beginning of this section, we can define a Weierstrass model based 
on the bundle $L$ on our compact K\"ahler manifold $B$. Since $c_1(L)=
c_1(B)$, we have constructed an elliptically fibred Calabi-Yau 4-fold
and hence a consistent F-theory model. The O7-plane, defined by the zero 
locus of $h$, is wrapped twice along the divisor since $h$ is a section of 
$L^{\otimes 2}$. At this point, we have already realized our local
Calabi-Yau as part of a compact type IIB model. It is intuitively clear
(although a better mathematical understanding would be desirable) that 
the O7-plane can be chosen in such a way that it does not interfere with the 
compact cycles of the original local Calabi-Yau. Indeed, the Calabi-Yau 
condition has been violated by making the original model compact. This 
violation is measured by the effective divisor associated with $L$. 
This divisor has therefore no need to pass through the region where the 
original compact cycles (relevant for local Calabi-Yau model building) are 
localized. We can even go one step further and separate the two O7-planes 
lying on top of the divisor of $L$. Subsequently, we can recombine them at 
possible intersection points, thereby arriving at a single smooth O-plane. 
Constructing the double cover of the base branched along this O-plane, we 
obtain a compact Calabi-Yau (without O-plane) the orientifolding of which 
takes us back to the above F-theory model~\cite{Sen:1997gv}. The compact 
Calabi-Yau without O-plane constructed in this way contains two copies of the 
original local Calabi-Yau model.

\section{General remarks on $K3$}\label{sec:K3}

In two complex dimensions there is, up to diffeomorphisms, just one compact 
Calabi-Yau manifold: $K3$. We will only collect the facts that we need; for 
a comprehensive review see e.g.~\cite{Aspinwall:1996mn}. 

The Hodge diamond of $K3$ is well known:
\be
\label{K3 hodge structure}
  {\arraycolsep=2pt
  \begin{array}{*{5}{c}}
    &&1&& \\ &0&&0& \\ 1&&20&&1. \\
    &0&&0& \\ &&1&&
  \end{array}}
\ee
The complex structure is measured by the periods $z_\alpha$ which are the 
integrals of the holomorphic 2-form $\Omega$ over integral 2-cycles. 
\be
z_{\alpha}\equiv\int_{\gamma_{\alpha}}\Omega=\int_{K3}
\eta_{\alpha}\wedge\Omega\equiv\eta_{\alpha}\cdot\Omega \ .
\ee
Here $\eta_{\alpha}$ are the Poincar\'e-dual 2-forms corresponding the 
2-cycles $\gamma_\alpha$. The real K\"ahler form $J$ can be decomposed in 
a basis of 2-forms in a similar way. Together, $\Omega$ and $J$ specify a 
point in the moduli space of K3. They have to fulfill the constraints
\be
\Omega \cdot \Omega = 0\quad,\qquad\quad J \cdot \Omega = 0\quad,\qquad\quad
\Omega \cdot \bar{\Omega}>0\quad,\qquad\quad J \cdot J >0\,.
\label{constraints}
\ee
Parameterizing $\Omega$ and $J$ by 3 real forms $x_i$, such that 
$\Omega = x_1 + \iu x_2$ and $J\sim x_3$, the constraints translate to
\be
\label{K3 moduli conditions I}
x_i \cdot x_j = 0 \qquad \textrm{for} \qquad i \neq j
\ee
and
\be
\label{K3 moduli conditions II}
x_1^2= x_2^2 = x_3^2 > 0 \ .
\ee \par
The symmetry between the three real 2-forms $x_i$ is 
related to the fact that $K3$ is a hyper-K\"ahler 
manifold: there is a whole $S^2$ of complex structures on $K3$.

The counting of oriented intersection numbers of 2-cycles
gives us a symmetric bilinear form on $H_2(K3,\mathbb{Z})$.
It can be shown~\cite{Aspinwall:1996mn} that with this 
natural scalar product, $H_2(K3,\mathbb{Z})$ is an even 
self-dual lattice of signature $(3,19)$. By the 
classification of even self-dual lattices we know that we 
may choose a basis for $H_2(K3,\mathbb{Z})$ such that the 
inner product forms the matrix
\be
\label{K3 moduli space inner product}
U \oplus U \oplus U \oplus -E_8 \oplus -E_8
\ee
where
\be
U = \left(
\begin{array}{cc}
0 & 1 \\ 1 & 0 \\
\end{array}
\right) \ ,
\ee
and $E_8$ denotes the Cartan matrix of $E_8$.
Choosing a point in the moduli space of $K3$ is 
now equivalent to choosing a space-like three-plane in $\mathbb{R}^{3,19}$ 
equipped with the inner product 
\eqref{K3 moduli space inner product}. This space-like 
three-plane is spanned by the three vectors $x_i$ fulfilling
the conditions~\eqref{K3 moduli conditions I} 
and~\eqref{K3 moduli conditions II}\footnote{
We
identify $H_2(K3,\mathbb{R})$ and $H^2(K3,\mathbb{R})$ here and below.
}. 

The Picard group, defined as
\be
\operatorname{Pic}(X)\equiv H^{1,1}(X)\cap H^{2}(X,\mathbb{Z}) \ ,
\ee
is given by the intersection of the lattice 
$H^{2}(X,\mathbb{Z})$ with the codimension-two surface 
orthogonal to the real and imaginary parts of $\Omega$. 
The dimension of $\operatorname{Pic}(X)$, also called Picard 
number, counts the number of algebraic curves and
vanishes for a generic $K3$ manifold.  

If we require $K3$ to admit an elliptic fibration, there are at least two 
algebraic curves embedded in $K3$ - the $T^2$ fiber and a section, the latter 
being equivalent to the base $\mathbb{CP}^1$. Thus the space orthogonal to 
the plane defining the complex structure has a two-dimensional intersection 
with the lattice $H^{2}(K3,\mathbb{Z})$, which fixes two complex structure 
moduli\footnote{
This 
behavior is a specialty of $K3$ and is related to the fact that,
contrary to higher-dimensional Calabi-Yau spaces, $h^{1,1}\neq b^{2}$.
}.
One can show that the two vectors in the lattice corresponding to the base 
and the fiber form one of the $U$ factors in~\eqref{K3 moduli space inner 
product}. Thus, $\Omega$ has to be orthogonal to the subspace corresponding 
to this $U$ factor. The precise position of $J$, which lies completely in this 
$U$ factor, is fixed by the requirement that the fibre volume goes to zero
in the F-theory limit. The only remaining freedom is in the complex 
structure, which is now defined by a space-like two-plane in 
$\mathbb{R}^{2,18}$ with the inner product
\be
\label{K3 complex structure moduli space inner product}
U \oplus U \oplus -E_8 \oplus -E_8 \ .
\ee
Any vector in the lattice of integral cycles of
an elliptically fibred $K3$ can now be written as
\begin{equation}
D=p^{i}e^{i}+p_{j}e_{j}+q_{I}E_{I},\label{H2}
\end{equation}
where $i,j$ run from one to two and $I,J$ from $1$ to $16$. 
The $p_{i}$ as well as the $p^{i}$ are all integers. The 
$E_{8}^{\oplus 2}$ lattice is spanned by $q_{I}$ fulfilling
$\sum_{I=1..8}q_{I}=2\mathbb{Z}$, 
$\sum_{I=9..16}q_{I}=2\mathbb{Z}$. In each of the two $E_{8}$
blocks, the coefficients furthermore have to be \emph{all} 
integer or \emph{all} half-integer~\cite{Polchinski:1998rr}.
The only non-vanishing inner products among the vectors 
in this expansion are
\begin{align}
E_{I}\cdot E_{J}=-\delta_{IJ} \hspace{1cm} e^{i}\cdot e_{j}=\delta^{i}_{j} \ .
\end{align}

There are $18$ complex structure deformations left in the elliptically 
fibred case: $\Omega$ may be expanded in twenty two-forms, 
which leads to $20$ complex coefficients. However, there is 
still the possibility of an arbitrary rescaling of $\Omega$ by one complex 
number, as well as the complex constraint $\Omega\cdot\Omega=0$, so that we 
find an $18$-dimensional complex structure moduli space. 

This number can easily be compared to the moduli of $K3$ considered 
as an elliptic fibration over $\mathbb{CP}^1$~\cite{Vafa:1996xn}, 
defined by the Weierstrass equation
\be
y^2=x^3+f_8(a,b) x+g_{12}(a,b) \label{weiersec4}
\ee
where $[a\!:\!b]$ are the homogeneous coordinates of $\mathbb{CP}^1$ 
and $f_8$ and $g_{12}$ are homogeneous polynomials of degree 8 
and 12 respectively. They are determined by $9+13=22$ parameters. 
There is an $SL(2,\mathbb{C})$ symmetry acting on the homogeneous 
coordinates of the base $\mathbb{CP}^1$ and an overall rescaling of 
\eqref{weiersec4}. This reduces the independent number of 
parameters to $18$ \cite{Vafa:1996xn}. From the perspective of 
F-theory compactified on $K3$, $17$ of these $18$ parameters describe
the locations of D7-branes and O-planes on the base of the fibration,
$\mathbb{CP}^{1}$. The remaining parameter describes the complex 
structure of the fiber and corresponds to the axiodilaton $\tau$. 
At the same time, these $18$ parameters describe the variation of 
the complex structure of the $K3$,
so that one can interpret D-brane moduli as complex structure 
moduli of an elliptically fibred higher-dimensional space.

\section{Singularities and the Weierstrass model.}\label{sgg}

As D7-branes are characterized by a degeneration of the 
elliptic fibre one may wonder whether the total space, in our
case $K3$, is still smooth. Its defining equation shows 
that $K3$ is generically smooth and only the fibration 
becomes singular. If the discriminant of the Weierstrass
model has multiple zeros, corresponding to placing multiple branes
on top of each other, the total space becomes singular as well.
In the case of elliptic $K3$ manifolds, the classification of singularities
matches the classification of the appearing gauge groups.

The zeros of the discriminant
\be
\Delta=4f^{3}+27g^{2}
\ee
determine the points of the base where the roots of the 
Weierstrass equation degenerate.
Let us examine this further. The Weierstrass equation 
\eqref{weiersec4} can always be written in the form\footnote{
Note 
that the term quadratic in $x$ is absent in the canonical 
form~\eqref{weiersec4}. This corresponds to choosing the origin of our 
coordinate system such that the three roots $x_i$ sum up to zero.
}
\be
y^{2}=(x-x_{1})(x-x_{2})(x-x_{3}) \ .
\ee
It is easy to show that in this notation
\be
\Delta=(x_{1}-x_{2})^{2}(x_{2}-x_{3})^{2}(x_{3}-x_{1})^{2} \ .
\label{singularities delta}
\ee
At a point where the fibre degenerates, two of the $x_i$ 
coincide so that, adjusting the normalization for 
convenience,~\eqref{weiersec4} reads locally
\be
y^{2}=(x-x_{0})^{2}\,.
\ee
By a change of variables this is equivalent to $xy=0$, representing an 
$A_{1}$ singularity of the fibre. To see what happens to the whole space we 
have to keep the dependence on the base coordinates. Let us deform away 
from the degenerate point by shifting $x_{0}\rightarrow x_{\pm} 
=x_{0}\pm\delta$. This means that now
\be
y^{2}=(x-x_{+})(x-x_{-})=(x-x_{0})^{2}+\delta^{2} \ .\label{y2ofdelta}
\ee
The quadratic difference of the now indegenerate roots is given by
\begin{equation}
(x_{+}-x_{-})^{2}=4\delta^{2} \ .
\end{equation}
Comparing this with~\eqref{singularities delta} and ignoring the slowly 
varying factor associated with the distant third root, we have 
\begin{equation}
\delta^2\sim\Delta\,.
\end{equation}
Since we also want to see what happens to the full space, we reintroduce the 
the dependence on the base coordinates, $\Delta=\Delta(a,b)$, and 
write~\eqref{y2ofdelta} as
\be
y^{2}=(x-x_{0})^{2}+\Delta(a,b) \ .
\ee

Without loss of generality we assume $a\neq 0$ and use $a$ as an 
inhomogeneous coordinate. Near the singularity, where $\Delta=(a-a_{0})^{n}$,
the Weierstrass model then reads
\be
y^{2}=(x-x_{0})^{2}+(a-a_{0})^{n}
\label{singularity structure}
\ee
which is clearly singular if $n$ is greater than one. By a 
change of variables this is again equivalent to
\be
yx=(a-a_{0})^{n} \ .
\ee
Thus, simple roots ($n=1$) of $\Delta$ do not lead to any singularity of the 
whole space, it is merely the fibration structure that becomes singular. 

We have seen that the $K3$ surface has a singularity when two or
more branes are on top of each other, whereas it is smooth 
when they are apart. If we move the two branes apart by 
perturbing the polynomials in the Weierstrass equation, we remove the 
singularity and blow up an exceptional 2-cycle. We will make this explicit 
in the following. The emerging cycle will subsequently be used to 
measure the distance between the two branes.

From \eqref{singularity structure} we see 
that the situation of two merging branes is described by
\begin{equation}
X\equiv\lbrace x,y,a \mid x^2+y^2+a^2=\epsilon\rbrace \ . 
\label{singularity ball}
\end{equation}
We have shifted $x$ and $a$ for simplicity. Here $a$ is an affine coordinate 
on the base and $\epsilon$ resolves the singularity by moving the two branes 
away from each other.

As the situation is somewhat analogous to the conifold case 
\cite{Candelas:1989ug}, we will perform a similar analysis:
We first note that $\epsilon$ can always be chosen to be real by redefining 
the coordinates. Next, we collect $x,y,a$ in a complex vector with real part 
$\xi$ and imaginary part $\eta$. The hypersurface~\eqref{singularity ball} 
may then be described by the two real equations 
\begin{equation}
\xi^2-\eta^2=\epsilon, \hspace{.5cm}
\xi\cdot\eta=0 \ .
\end{equation}
We can understand the topology of X by considering its intersection with 
a set of 5-spheres in $\mathbb{R}^6$ given by $\xi^2+\eta^2=t$, 
$t>\epsilon$ :
\begin{equation}
\xi^2=\frac{t}{2}+\frac{\epsilon}{2}, \hspace{.5cm}
\eta^2=\frac{t}{2}-\frac{\epsilon}{2}, \hspace{.5cm}
\xi\cdot\eta=0 \ . \label{s2s1}
\end{equation}
If we assume for a moment that $\epsilon=0$, the
equations above describe two $S^2$s of equal size for
every $t$ that are subject to an extra constraint. 
If we take the first $S^2$ to be unconstraint, 
the second and third equation describe the intersection of 
another $S^2$ with a hyperplane. Thus we have an $S^1$ 
bundle over $S^2$ for every finite $t$. This bundle shrinks
to zero size when $t$ approaches zero so that we reach the 
tip of the cone. Furthermore, the bundle is clearly non-trivial since 
the hyperplane intersecting the second $S^2$ rotates as one moves along the 
first $S^2$. 

Let us now allow for a non-zero $\epsilon$, so that $X$ is no 
longer singular. The fibre $S^1$ still shrinks to zero size at 
$t=\epsilon$, but the base $S^2$ remains at a finite
size. This is the 2-cycle that emerged when resolving the singularity.
We now want to show that the $S^2$ at the tip of the resolved 
cone is indeed a non-trivial cycle. This is equivalent to 
showing that the bundle that is present for $t>\epsilon$ 
does not have a global section, i.e., the $S^2$ at $t=\epsilon$ cannot 
be moved to larger values of $t$ as a whole. Consider a small deformation 
of this $S^2$, parameterized by a real function $f(\xi)$:
\begin{equation}
\xi^2=\epsilon+f(\xi) \ .
\end{equation}
This means that we have moved the $S^2$ at $t=\epsilon$ in the $t$-direction 
by $2f(\xi)$, i.e.
\begin{equation}
\frac{t}{2}=f(\xi)+\frac{\epsilon}{2} \ .
\end{equation}
The equations for the $S^1$ fibration over this deformed $S^2$ 
read
\begin{align}
\eta^2&=f(\xi) \label{etasq} \\ 
\eta\cdot\xi&=0\label{etaf} \ .
\end{align}
Deforming the $S^2$ means choosing a continuous function $\eta(\xi)$ subject 
to these equations. Since the set of all planes defined by~\eqref{etaf} 
can be viewed as the tangent bundle of an $S^2$, we can view  $\eta(\xi)$
as a vector field on $S^2$. As we know that such vector
fields have to vanish in at least two points, we conclude that $f(\xi)$
has to vanish for two values of $\xi$. We do not only learn that
the $S^2$ at the tip of the cone cannot be moved away, but also
that the modulus of its self-intersection number is two\footnote{
Note 
that this viewpoint also works for the classic conifold example~\cite{
Candelas:1989ug}, which is a cone whose base is an $S^2$ bundle over $S^3$. 
As the Euler characteristic of $S^3$ vanishes, the same arguments as before 
give the well-known result that the $S^3$ at the tip of the resolved conifold 
can be entirely moved into the base.
}.
This number is expected, as one can show that
any cycle of $K3$ with the topology of a sphere
has self-intersection number minus two~\cite{Aspinwall:1996mn}.

A similar analysis can be carried out for other
types of singularities, which are determined by the orders 
with which $f$, $g$ and $\Delta$ vanish~\cite{Morrison:1996na}. 
For the whole $K3$, we can use the ordinary ADE classification 
of the arising quotient singularities~\cite{Aspinwall:1996mn}, which is
equivalent to the classification of simply laced Lie algebras. 
The intersection pattern of the cycles that emerge in
resolving the singularity is precisely the Dynkin diagram of the
corresponding Lie algebra.
This `accidental' match is of course expected from the F-theory point
of view which tells us that  singularities are associated
with gauge groups.
The correspondence between the singularity 
type of the whole space and the singularity type of the 
fibre is more complicated when F-theory is
compactified on a manifold with more than two complex dimensions
~\cite{Aspinwall:1995xy,Aspinwall:1996nk,Aspinwall:2000kf,Bershadsky:1996nh}.
This can also be anticipated from the fact that
gauge groups which are not simply laced appear only in IIB 
orientifolds with more than one complex dimensions.

\section{Geometric picture of the moduli space}\label{gp}

In this section we want to gain a more intuitive understanding of the cycles
that are responsible for the brane movement. For this we picture the 
elliptically fibred K3 locally as the complex plane (in which branes are 
sitting) to which a torus has been attached at every point. We will construct 
the relevant 2-cycles geometrically. We have already seen that the cycles in 
question shrink to zero size when we move the branes on top of each other, 
so that these cycles should be correlated with the distance between the
branes. Remember that D7-branes have a non-trivial monodromy acting on the 
complex structure of the fibre as $\tau \to \tau +1$, which has 
\be
T = \left( \begin{array}{cc}
1 & 1 \\
0 & 1 \\
\end{array}
\right) 
\label{monodromy matrix T}
\ee
as its corresponding $SL(2,\mathbb{Z})$ matrix.
Similarly, O7-planes have a monodromy of $-T^4$, where 
the minus sign indicates an involution of the torus,
meaning that the complex coordinate $z$ of 
the torus goes to $-z$. Thus 1-cycles 
in the fibre change orientation when they are moved around 
an O-plane. 

\begin{figure}[!ht]
\begin{center}
\includegraphics[height=3.5cm,]{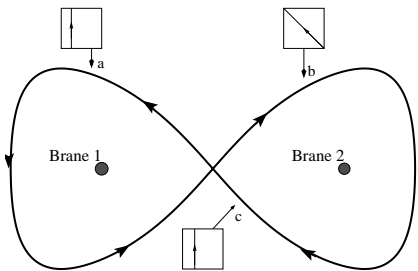}\caption{\textit{The 
cycle that measures the distance between
two D-branes. Starting with a cycle in the $(0,1)$ direction 
of the fibre torus at point $a$, this cycle is tilted to 
$(-1,1)$ at $b$. Because we surround the second brane in the  
opposite way, the cycle in the fibre is untilted again so it 
can close with the one we started from.}}
\label{Cyclebetweentwobranes}
\end{center}
\end{figure}

If we want to describe a 2-cycle between two D-branes, 
it is clear that it must have one leg in the base and one
in the fibre to be distinct from the 2-cycles describing 
the fibre and the base. Now consider the 1-cycle being 
vertically stretched in the torus\footnote{Of course this 
notion depends on the $SL(2,\mathbb{Z})$-frame we consider, 
but anyway we construct the cycle in the frame where the 
branes are D-branes. In the end the constructed cycle will 
be independent of the choice of frame.}. If we transport 
this cycle once around a D-brane and come back to the same 
point, this cycle becomes diagonally stretched because of 
the T-monodromy. If we then encircle another brane in the 
opposite direction, the 1-cycle returns to its original form, so that it 
can be identified with the original 1-cycle. 
This way to construct a closed 2-cycle 
was already mentioned in~\cite{Sen:1996vd}. 
This 2-cycle cannot be contracted to a point since it 
cannot cross the brane positions because of the monodromy 
in the fibre. The form of the 2-cycle is illustrated in 
Fig.~\ref{Cyclebetweentwobranes}. We emphasize that to get a non-trivial 
cycle, its part in the fibre torus has to have a vertical component.

\begin{figure}[!ht]
\begin{center}
\includegraphics[height=3.5cm,]{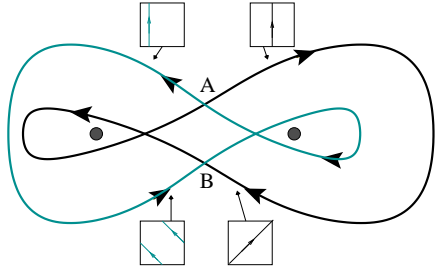}\caption{\textit{The 
self-intersection number of a cycle between two D-branes. 
As shown in the picture, we may choose the fibre part of both cycles to be 
$(0,1)$ at $A$, so that they do not intersect at this point. At $B$ 
however, one of the two is tilted to $(1,1)$, whereas the other has undergone 
a monodromy transforming it to $(-1,1)$. Thus the two surfaces
meet twice in point $B$.}}\label{Cyclesbetweentwobranesintersecting}
\end{center}
\end{figure}

\begin{figure}[!ht]
\begin{center}
\includegraphics[height=3.5cm,]{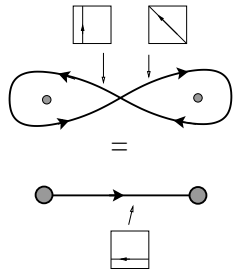}\caption{\textit{The loop
between two D-branes can be collapsed to a line by 
pulling it onto the D-branes and annihilating the
vertical components in the fibre. All that remains
is a cycle which goes from one brane to the other
while staying horizontal
in the fibre all the time.}}
\label{looptoline}
\end{center}
\end{figure}

Next we want to compute the self-intersection number.
In order to do this, we consider a homologous cycle and 
compute the number of intersections with the original one.
By following the way the fibre part of the cycle evolves, one 
finds that the resulting number is minus two 
(see Fig.~\ref{Cyclesbetweentwobranesintersecting}). The minus 
sign arises from the orientation. This is precisely what we expected 
from the previous analysis. To see the topology of the cycle more clearly,
it is useful to combine the two lines in the base stretching between the 
branes to a single line. This is shown in Fig.~\ref{looptoline}. 
The component of the cycle in the fibre is then an $S^1$
that wraps the fibre in the horizontal direction, so that 
it shrinks to a point at the brane positions. 
Thus it is topologically a sphere, which fits 
with the self-intersection number of $-2$ and the
discussion of the previous section.

Now we want to determine the intersection number between 
different cycles and consider a situation with 
three D7-branes. There is one cycle between
the first two branes and one between the second and the third, 
each having self-intersection number $-2$. From 
Fig.~5 it should be clear
that they intersect exactly once.
If one now compares the way they intersect to the
figure that was used to determine the self-intersection number,
one sees that the two surfaces meet with one direction reversed, 
hence the orientation differs and we see that the mutual intersection 
number is $+1$. Thus we have shown that the intersection 
matrix of the $N-1$ independent cycles between $N$ D-branes is 
minus the Cartan Matrix of $SU(N)$.

\begin{figure}[!ht]
\begin{center}
\label{3branes}
\includegraphics[height=3.5cm,]{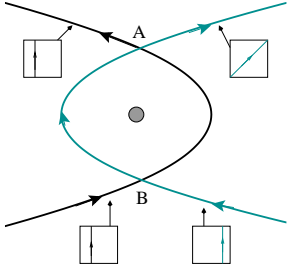}\caption{\textit{Mutual 
intersection of two cycles.
Start by taking both cycles to have fibre part $(0,1)$ at $B$. The fact
that we closed a circle around the D-brane tells us that one of the two
has been tilted by one unit at $A$. Thus they meet precisely once.}}

\end{center}

\end{figure}

\begin{figure}[!ht]
\begin{center}
\includegraphics[height=3cm,]{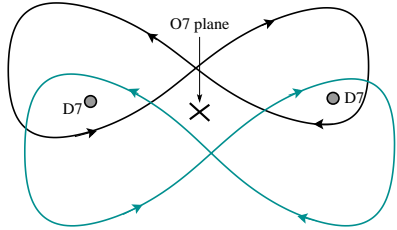}\caption{\textit{Cycles that measure
how D-branes can be pulled onto O-planes.}}\label{DandO}
\end{center}
\end{figure}

We now want to analyze the cycles that arise in the
presence of an O7-plane. Two D7-branes in the vicinity of an
O7-plane can be linked by the type of cycle considered above. 
However, there are now two ways to connect the D-branes with each other: 
we can pass the O-plane on two different sides, as shown in Fig.~\ref{DandO}.
By the same argument as before, each of these cycles has
self-intersection number $-2$. To get their
mutual intersection number, it is important to
remember the monodromy of the O-plane, which contains an involution of the torus 
fiber. Thus, the intersection on the right and the intersection on the left, 
which differ by a loop around the O-plane, have opposite sign. As a result, 
the overall intersection number vanishes.

\begin{figure}[ht!]
\begin{center}
\includegraphics[height=2cm,]{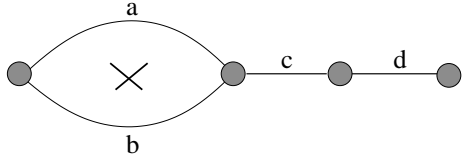}\caption{\textit{Four D-branes and
an O-plane. The D-branes are displayed as circles and
the O-plane as a cross. To simplify the picture we have drawn
lines instead of loops.}}\label{four D and and an O}
\end{center}
\end{figure}

Now we have all the building blocks needed to discuss the 
gauge enhancement in the orientifold limit, which is $SO(8)^4$. 
We should find an $SO(8)$ for each O7-plane with four D7-branes on top of 
it. The cycles that are blown up when the four D7-branes move away from the 
O7-plane are shown in Fig.~\ref{four D and and an O}. It is clear from the 
previous discussion that all cycles have self-intersection number $-2$ and 
cycle $c$ intersects every other cycle precisely once. Thus, collecting the 
four cycles in a vector $(a,b,c,d)$, we find the intersection form
\begin{equation}
D_4=\left(\begin{array}{cccc} -2 & 0 & 1 & 0\\
			   0 & -2 & 1 &0  \\
			   1 &  1 & -2 & 1 \\
			   0 & 0  & 1 & -2 \end{array}\right) 
\label{D4matrix} \ ,
\end{equation}
which is minus the Cartan matrix of $SO(8)$. It is also easy to
see that collapsing only some of the four cycles yields minus the Cartan
matrices of the appropriate smaller gauge enhancements.

\section{Duality to M-theory and heterotic $E_8 \times E_8$}\label{het}

In this section we give a brief review of the dualities between F-theory and 
M-theory, as well as between F-theory and the $E_8 \times E_8$ heterotic 
string. We focus on the points relevant to the discussion in this paper. 

As M-theory compactified on $S^1$ is dual to type IIA in 10 dimensions, we 
can relate M-theory to type IIB by compactifying on a further $S^1$ and 
applying T-duality. Thus, M-theory on $T^2$ corresponds to type IIB on 
$S^1$. The complexified type-IIB coupling constant is given by the complex 
structure of the torus. Furthermore, taking the torus volume to zero 
corresponds to sending the $S^1$ radius on the type IIB side to infinity. In 
other words, M-theory on $T^2$ with vanishing volume gives type IIB in 10 
dimensions. One may think of this as a $T^2$ compactification of a 12d theory, 
which we take as our working definition of F-theory. Considering the 
compactification of M-theory on an elliptically fibred manifold $Y$ and using 
the above argument for every fibre, we arrive at type IIB on the base space. 
This can be re-expressed as an F-theory compactification on $Y$. 

One can now study the stabilization of D7-branes through fluxes by 
investigating the stabilization of a (complex) fourfold $Y$ on the
M-theory side and mapping the geometry to the D-brane 
positions~\cite{Gorlich:2004qm, Lust:2005bd}. One can also translate the 
4-form fluxes of M-theory to 3- and 2-form fluxes in the IIB picture. 
In particular, we can consider 4-cycles built from a 2-cycle stretched 
between two D7-branes (see previous section) and a 2-cycle of the D7-brane. 
We then expect 2-form flux on D7-branes to arise from M-theory 4-form flux
on such cycles. This fits nicely with the fact that 2-form flux on 
D7-branes T-dualizes to an angle between two intersecting D6-branes,
so that it is only defined relative to the flux on another D7-brane.
Note that being in the F-theory limit (i.e. ensuring that $Y$ is elliptically 
fibred and the fiber volume vanishes) also has to be realized by an 
appropriate flux choice on the M-theory side.

The foundation of the duality between F-theory and the $E_8\times E_8$ 
heterotic string is the duality between M-theory compactified on $K3$ and the 
heterotic string compactified on $T^3$~\cite{Witten:1995ex}. In the limit in 
which the fibre of $K3$ shrinks (the F-theory limit), one $S^1$ 
decompactifies so that we end up with heterotic $E_8\times E_8$ on $T^2$
\cite{Vafa:1996xn, Vafa:1997pm, LopesCardoso:1996hq, 
Lerche:1998nx, Lerche:1999de}.

In this duality the complex structure of the base and the fibre of $K3$ are 
mapped to the complex and K\"ahler structure moduli of the $T^2$ on the 
heterotic side. The precise relation between these parameters has been 
worked out in~\cite{LopesCardoso:1996hq}. The volume of the base
of $K3$ corresponds to the coupling of the heterotic theory. The breaking 
of $E_8\times E_8$ that is achieved by Wilson lines on the heterotic
side appears in the form of deformations of the Weierstrass equation away 
from the $E_8\times E_8$ singularity on the F-theory side. This is 
equivalent to deforming the complex structure of $K3$.

Let us examine this point in more detail:
we can parameterize the complex structure of $K3$
at the $E_8\times E_8$ point by
\be
\Omega_{E_8\times E_8}= 
e^1-\tilde{U}\tilde{S}e_1+\tilde{U}e^2+\tilde{S} e_2 \ .
\label{Omega at E8 point}
\ee
We have set the first coefficient to one by a global rescaling 
and used the fact that $\Omega\cdot\Omega=0$. 
We can now add a component in the direction of the $E_8$ factors 
to~\eqref{Omega at E8 point} in order to break the gauge group.

In the heterotic theory, 
the Wilson lines are two real vectors that are labelled
by their direction in $T^2$: $W^1_I$ and $W^2_I$. Let us 
normalize these Wilson lines such that the associated 
gauge-theoretical twist is \be
P = \e^{-2 \pi \iu W^1_I E_I} \ ,
\ee
with the action
\be
E_\gamma \to P E_\gamma P^{-1} = \e^{-2 \pi \iu \gamma_I 
W^1_I } E_\gamma \ ,
\ee
and analogously for $W^2_I$. This means that, whenever there
is a root vector $\gamma$ such that $\gamma \cdot W^1$ and
$\gamma \cdot W^2$ are integers, the corresponding root 
survives the Wilson line breaking.

Let us define $W_I E_I = W^1_I E_I + \tilde{U}W^2_I E_I$ 
and write the complex structure at a general point in moduli 
space as

\begin{equation}
\Omega=e^1+\tilde{U}e^2+\tilde{S}e_2-\left(\tilde{U}\tilde{S}
+\frac{1}{2}(W)^2\right)e_1+W_I E_I \ . 	     \label{omegagen}
\end{equation}

Here we denote $(W^1_I E_I + \tilde{U}W^2_I E_I)^2$ simply by 
$(W)^2$. Note that $\Omega\cdot\Omega=0$ holds for all values 
of the parameters. If we find a surviving root $\gamma=\gamma_I E_I$
in the $E_8\times E_8$ lattice, its inner product with the 
complex structure is $\Omega\cdot\gamma=-n-\tilde{U}m$. This 
means there exists a cycle $\gamma'\equiv\gamma_I E_I+ne_1+me_2$ with the 
property $\Omega\cdot\gamma'=0$, implying that this cycle has shrunk to
zero size. Thus, extra massless states are present and a gauge enhancement 
arises, which is precisely what one expects from a surviving root on the 
heterotic side. We can conclude that, by (\ref{omegagen}), we have 
consistently identified the properly normalized Wilson lines $W^1$ and $W^2$ 
of the heterotic description with the appropriate degrees of freedom of the 
complex structure of K3 on the F-theory side.

\section{The $SO(8)^4$ singularity of $K3$}\label{so8}

If we describe the possible deformations of $K3$ by 
deformations of the defining functions of the Weierstrass 
model, $f$ and $g$, we see that singularities arise 
at specific points in moduli space. 
We can also describe 
the deformations of K3 as deformations of the complex 
structure, $\Omega$. In this description we can also reach points 
in the moduli space where the K3 is singular. By comparing 
the singularities, we can map special values of the complex 
structure moduli to a special form of the Weierstrass model.
This enables us to describe the positions of the D-branes and
O-planes, which are explicit in the polynomial description,
by the values of the complex structure moduli of $K3$.

In the language of complex structure deformation, the 
K3 becomes singular if we can find a root, i.e. an 
element of $H^{2}(\mathbb{Z})$ with self intersection 
$-2$ that is orthogonal to both $\Omega$ and $J$. If 
we now consider the inner product on the sublattice 
spanned by such roots, we find minus the Cartan Matrix 
of some $ADE$ group. This structure tells us the kind of singularity 
that has emerged~\cite{Aspinwall:1996mn,Gorlich:2004qm}.

We now want to discuss the cycles that shrink to produce an 
$SO(8)^4$ singularity. For this we have to find a change of 
basis such that the $D_4^{\oplus 4}$ in the inner product of 
the lattice $H^{2}(K3,\mathbb{Z})$ becomes obvious. The well-known Wilson 
lines breaking $E_8 \times E_8$ down to $SO(8)^4$ are
\begin{align}
W^1=\left(0^4,\frac{1}{2}^4,0^4,\frac{1}{2}^4\right)\hspace{1cm}
W^2=\left(1,0^7,1,0^7\right) \ .
\label{Wilson lines SO8}
\end{align}
$W^2$ breaks $E_8 \times E_8$ down to $SO(16) \times SO(16)$, 
$W^1$ breaks this further down to $SO(8)^4$.
Inserting this in~\eqref{omegagen}, we find $\Omega$ at the $SO(8)^4$ point:
\begin{align}
\Omega_{SO(8)^4}=e^1+\tilde{U}e^2+\tilde{S}e_2-\left(\tilde{U}\tilde{S}
-1-\tilde{U}^2\right)e_1+W_I E_I \ .
\label{omegaso8}
\end{align}
Note that setting $\tilde{S}=2\iu$ and $\tilde{U}=\iu$ 
reproduces the complex structure given 
in~\cite{Gorlich:2004qm}.

The lattice vectors orthogonal to $\Omega_{SO(8)^4}$ span the lattice 
$D_4^{\oplus 4}$. Using the expansion~(\ref{H2}), their coefficients have to 
satisfy
\begin{equation}
-\left(\tilde{U}\tilde{S}-1-\tilde{U}^2\right)p^1+p_1-W^1_{I}q_{I}+
\tilde{S}p^2+\tilde{U}(p_2-W^2_{I}q_{I})=0 \ .
\end{equation}
As we know that these lattice vectors must be orthogonal
to the complex structure for every value of $\tilde{U}$ and
$\tilde{S}$, we find the conditions
\begin{align}
p^1=0 &\hspace{1cm} p_1-W^1_{I}q_{I}=0 \nonumber \\
p^2=0 &\hspace{1cm} p_2-W^2_{I}q_{I}=0 \ .
\end{align}
These equations are solved by the following four groups of 
lattice vectors:

\vspace{.5cm}

\begin{center}
\begin{tabular}[h]{c|c|c|c|c}
& $A$&$B$&$C$&$D$\\
\hline
1&$E_{7}-E_{8}$&$-E_{15}+E_{16}$&
$-e_2-E_{1}+E_{2}$&$e_2+E_{9}-E_{10}$\\
2&$E_{6}-E_{7}$&$-E_{14}+E_{15}$ &
$-E_{2}+E_3$&$E_{10}-E_{11}$\\
3&$-e_1-E_{5}-E_{6}$&$e_1+E_{13}+E_{14}$&
$-E_{3}+E_{4}$&$E_{11}-E_{12}$\\
4&$E_{5}-E_{6}$&$-E_{13}+E_{14}$&
$-E_{3}-E_{4}$&$E_{11}+E_{12}$       

\end{tabular} 
\vspace{-1.7cm}
\begin{equation}
\label{AtoD}
\end{equation}
\end{center}

\vspace{1.5cm}

It is not hard to see that there are no mutual intersections between the 
four groups, and that the intersections within each group are given by the 
$D_4$ matrix~\eqref{D4matrix}. This serves as an explicit check 
that~(\ref{omegaso8}) is indeed the correct holomorphic two-form of 
$K3$ at the $SO(8)^4$ point.

\begin{figure}[ht!]
\begin{center}
\includegraphics[height=2cm,]{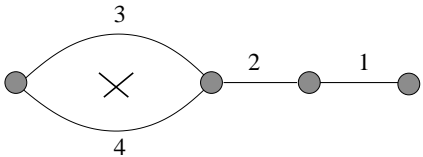}\caption{\textit{The assignment between 
the geometrically constructed cycles between branes and the cycles of the 
table in the text. Note that the distribution of the cycles 1,3 and 4 is 
ambiguous.\label{1block}}}
\end{center}
\end{figure}

It should be clear that one can choose different linear combinations 
of the basis vectors in each block that still have the same inner 
product. This only means we can describe the positions of the D-branes by 
a different combination of cycles, which are of course linearly dependent on 
the cycles we have chosen before and span the same lattice.
We can make an assignments between the cycles in the table and the cycles 
constructed geometrically as shown in Fig.~\ref{1block}.

\begin{figure}[ht!]
\begin{center}
\includegraphics[height=5cm,]{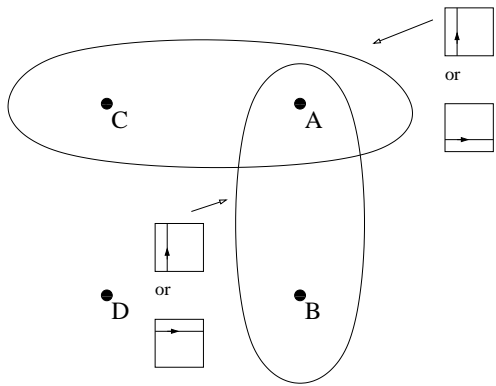}\caption{\textit{The
torus cycles have to encircle two blocks to ensure trivial monodromy 
along the basis part of the cycles. Note that the cycles which are
orthogonal in the base and in the fibre intersect twice because of the
orientation change introduced in going around the O-plane.\label{fourt}}}
\end{center}
\end{figure}

When we are at the $SO(8)^4$ point, where $16$ of the $20$ cycles of $K3$ 
have shrunk, the only remaining degrees of freedom are the deformations of 
the remaining $T^2/Z_2\times T^2$. The four cycles describing these 
deformations have to be orthogonal to all of the $16$ brane cycles. There 
are four cycles satisfying this requirement,
\begin{align}
e^1+W_{I}^1E_{I} &\hspace{1cm} e^2+W^2_IE_I \nonumber \\
e_1 & \hspace{1cm} e_2 \ ,		\label{orthomega}
\end{align}
and the torus cycles must be linear combinations of them.
From the fibration perspective, the torus cycles are the cycles 
encircling two blocks (and thus two O-planes), so that the monodromy along 
the base part of those cycles is trivial. They can be either horizontal or 
vertical in the fibre, giving the four possibilities displayed in 
Fig.~\ref{fourt}. Note that all of them have self-intersection zero and only 
those that wrap both fibre and base in different directions intersect 
twice. 

To find out which linear combination of the forms in (\ref{orthomega}) 
gives which torus cycle, we will consider a point in moduli space where 
the gauge symmetry is enhanced from $SO(8)^4$ to $SO(16)^2$. At this point, 
there are $16$ integral cycles orthogonal to $\Omega$ the intersection matrix 
of which is minus the Cartan matrix of $SO(16)^2$. Furthermore, we know 
that this situation corresponds to moving all D-branes onto two O7-planes. 
We will achieve this leaving two of the four blocks untouched, while 
moving the D-branes from the other blocks onto them. This means that we blow 
up one of the cycles in each of the blocks that are moved, while collapsing 
two new cycles that sit in between the blocks. Doing this we find three 
independent linear combinations of the cycles in~(\ref{orthomega}) that do 
not intersect any of the cycles that are shrunk. 

Before explicitly performing this computation, we choose a 
new basis that is equivalent to \eqref{orthomega}:
\begin{align}
\alpha\equiv2\left(e^1+e_1+W^1_IE_I\right) 
&\hspace{1cm}\beta\equiv2(e^2+e_2+W^2_IE_I)\nonumber \\
e_1 &\hspace{1cm} e_2 \ .	\label{basistcyc}
\end{align}
In this basis we can write $\Omega$ at the $SO(8)^4$ point as
\begin{align}
\Omega_{SO(8)^4}=\frac{1}{2}\left(\alpha+Ue_2+S\beta-USe_1\right) \ . 
\label{omegaUS}
\end{align}
We also have switched to a new parameterization in terms of 
$U$ and $S$. They will turn out to be the complex structures of the 
base and the fibre torus.

\begin{figure}[ht!]
\begin{center}
\includegraphics[height=2cm,]{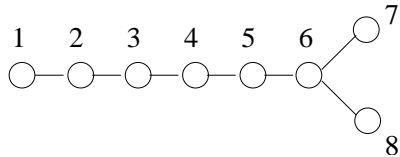}\caption{\textit{The Dynkin diagram of 
$SO(16)$.}}
\label{dynkin16}
\end{center}
\end{figure}

Let us now go to the $SO(16)^2$ point by setting $W^1_I=0$ in 
(\ref{omegagen}). After switching again from $\tilde{S}$ and $\tilde{U}$ to 
$S=\tilde{U}$ and $U/2=\tilde{S}-\tilde{U}$, we find
\begin{equation}
\Omega_{SO(16)^2}=e^1+\frac{U}{2}e_2+\frac{S}{2}\beta-\frac{US}{2}e_1 \ .  \label{omegaso16} 
\end{equation}
The $16$ integral cycles that are orthogonal to $\Omega$ are:
\begin{center}
\begin{tabular}[h]{c|c|c}
 & $E$&$F$\\  \hline
1&$-e_2-E_1+E_2$&$e_2+E_9-E_{10}$\\
2&$-E_2+E_3$&$E_{10}-E_{11}$\\
3&$-E_3+E_4$&$E_{11}-E_{12}$\\
4&$-E_4+E_5$&$E_{12}-E_{13}$\\
5&$-E_5+E_6$&$E_{13}-E_{14}$\\
6&$-E_6+E_7$&$E_{14}-E_{15}$\\
7&$-E_7+E_8$&$E_{15}-E_{16}$\\
8&$E_7+E_8$&$E_{15}+E_{16}$\\

\end{tabular}
\vspace{-2.8cm}
\begin{equation}
\label{EandF}
\end{equation}

\end{center}
\vspace{2cm}
They are labelled as shown in Fig.~\ref{dynkin16}. Note that this means that 
we have moved block $A$ onto block $C$ and block $B$ onto block $D$.

One can check that, out of the basis displayed 
in (\ref{basistcyc}), only $\alpha$ has a 
non-vanishing intersection with some of the $16$ forms
above, whereas all of them are orthogonal to
$e_1,e_2,\beta$. Thus, $\alpha$ is contained in the cycle that
wraps the fibre in vertical direction and passes in between $A,B$ and $C,D$
(cf. Fig.~\ref{fourt}). Furthermore, the non-zero intersection
between $e_1$ and $\alpha$ tells us that $e_1$ wraps the fibre 
horizontally (for this argument we used $e_1\cdot e_2=e_1\cdot\beta=0$).
Given these observations, it is natural to identify the four cycles 
(\ref{basistcyc}) with the four cycles displayed in Fig.~\ref{fourt}. 
More specifically, we now know that $\alpha$ is vertical in the fibre and 
passes in between $A,B$ and $C,D$, while $e_1$ is horizontal in the fibre 
and passes in between $A,C$ and $B,D$. The four cycles characterize 
the shape of $T^2/Z_2\times T^2$. Other possible assignments between 
the cycles of (\ref{basistcyc}) and those displayed in Fig.~\ref{fourt} 
correspond to reparameterizations of the tori and are therefore 
equivalent to our choice. 

\begin{figure}[ht!]
\begin{center}
\includegraphics[height=3cm,]{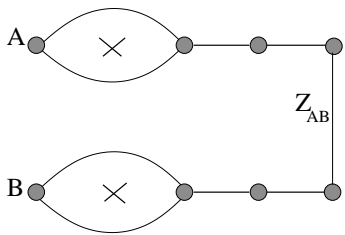}\caption{\textit{The cycles
connecting the blocks. Note that all cycles in this
picture are built in the same way and thus all
lie horizontally in the fibre.\label{ZAB}}}
\end{center}
\end{figure}

We now have to assign the cycles $e_2$ and $\beta$ to the two remaining 
cycles of Fig.~\ref{fourt}. For this purpose, we will explicitly construct 
the cycles $Z_{XY}$ between the four $SO(8)$ blocks. Since they can be 
drawn in the same way as the cycles between the D-branes, 
cf.~Fig.~\ref{ZAB}, we see that all of them must have self-intersection 
number $-2$. We also know that their mutual intersections should be 
$Z_{XY}\cdot Z_{YZ}=1$. From what we have learned so far, all of them should 
be either orthogonal to $\beta$ and $e_1$ or orthogonal to $e_2$ and $e_1$. 
It is easy to check that the first case is realized by:
\begin{align}
Z_{AC}=E_{8}-E_1-e_2&\hspace{1cm}Z_{AB}=-(e_2+e^2)-E_1+E_8-E_9+E_{16}+e_1 
\nonumber \\
Z_{DB}=E_{16}-E_{9}-e_2&\hspace{1cm}Z_{CD}=e_2-e^2 \ .
\end{align}
One can show that the second case is not possible. This can be seen 
from the following argument:

If we can find $Z$-cycles that are orthogonal to $e_1$ and $e_2$,
we can decompose them as
\be
Z_{XY}=q_I E_I \ .
\ee
Note that the $e_i$ are now responsible for making the $Z$-cycles wind 
around the base torus, so that we do not loose any generality by omitting 
them in the decomposition above.
Because of the constraint $\sum q_{I}=2\mathbb{Z}$, we can only have
$Z_{AB}$ intersecting one of the cycles in block $A$ by putting
$q_I=\pm \frac{1}{2}$ appropriately for $I=5..8$. By the structure
of the lattice, \eqref{H2}, this forces 
us to also set $q_I=\pm \frac{1}{2}$ for $I=1..4$. It is clear that
this will also make this cycle intersect with one of the cycles in block $C$, 
contradicting one of its defining properties. This means that we simply 
cannot construct the $Z$-cycles to be all orthogonal to $e_1$ and $e_2$. 

We can now use the intersections of the $Z$-cycles with
the complex structure at the $SO(8)^4$ point,
eq.(\ref{omegaUS}),
to measure their length and consistently distribute 
the four blocks on the pillow. We find that
\begin{align}
Z_{AB}\cdot\Omega_{SO(8)^4}=-U/2 & \hspace{1cm}
Z_{AC}\cdot\Omega_{SO(8)^4}=-\frac{1}{2} \nonumber \\
Z_{CD}\cdot\Omega_{SO(8)^4}=-U/2 & \hspace{1cm}
Z_{DB}\cdot\Omega_{SO(8)^4}=-\frac{1}{2} \ .
\end{align}

\begin{figure}[!ht]
\begin{center}
\includegraphics[height=7cm,]{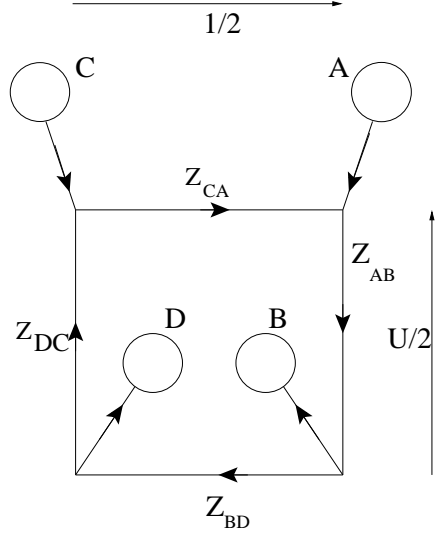}\caption{\textit{A schematic picture 
of the cycles between the blocks. We have drawn arrows to indicate the 
different orientations. Note that the $Z$-cycles have intersection
$+1$ with the cycle they come from and $-1$ with the cycle they go to.
Note that they sum up to a torus cycle, $e_1$, telling us
which blocks are encircled.}\label{zcycles}}
\end{center}
\end{figure}

\begin{figure}[!ht]
\begin{center}
\includegraphics[height=8cm,]{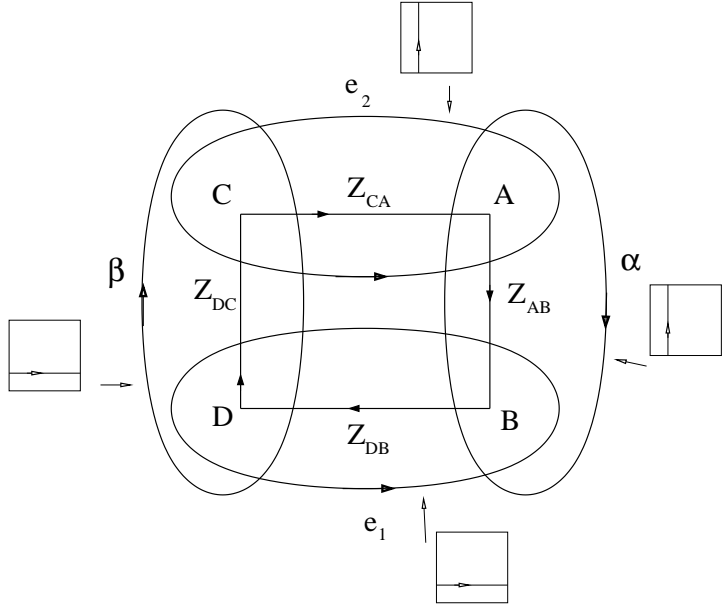}\caption{\textit{The distribution
of the torus cycles.}\label{tcyc}}
\end{center}
\end{figure}

It is clear that we can add the two orthogonal
cycles $e_1$ and $\beta$ to the $Z$-cycles,
$Z\rightarrow Z+ne_1+m\beta$, without destroying their
mutual intersection pattern. However, this changes their length
by $n+Um$. This means that we can make the $Z$-cycles wind around the 
pillow $n$ times in the real and
$m$ times in the imaginary direction. Calling the real 
direction of the base (fibre) $x$ ($x'$) and the imaginary direction
of the base (fibre) $y$ ($y'$), we can now make the
identifications:
\begin{align}
e_1 \hspace{1ex}\mbox{winds around}\hspace{1ex}
x\hspace{1ex} \mbox{and}\hspace{1ex} x' \nonumber \\
e_2 \hspace{1ex}\mbox{winds around}\hspace{1ex}
x \hspace{1ex}\mbox{and}\hspace{1ex} y'\nonumber \\
\alpha \hspace{1ex}\mbox{winds around}\hspace{1ex}
y \hspace{1ex}\mbox{and}\hspace{1ex} y'\,\nonumber \\
\beta \hspace{1ex}\mbox{winds around}\hspace{1ex}
y \hspace{1ex}\mbox{and}\hspace{1ex} x'.
\end{align}
Alternatively one can find the positions of $e_2$ and
$\alpha$ by computing their intersections with
the $Z$-cycles:
\begin{align}
e_2\cdot Z_{CD}=e_2\cdot Z_{AB}&=-1,\hspace{1cm}
e_2\cdot Z_{AC}=e_2\cdot Z_{DB}=0\,,\nonumber \\
\alpha\cdot Z_{AC}=\alpha\cdot Z_{DB}&=-1,\hspace{1cm}
\alpha\cdot Z_{CD}=\alpha\cdot Z_{AB}=0 \ .
\end{align}
Note that these intersections change consistently when we let 
$Z\rightarrow Z+ne_1+m\beta$. We display the distribution
of the torus cycles in Fig.~\ref{tcyc}.

At the orientifold point we can write the complex structure
of $T^2/Z_2\times T^2$ as
\begin{align}
\Omega_{T^2/Z_2\times T^2}=&\left(dx+Udy\right)\wedge\left(dx'+Sdy'\right) 
\nonumber \\
=& \hspace{1ex}dx\wedge dx'+Sdx\wedge dy'+Udy\wedge dx'+SUdy\wedge dy'\ .
\end{align}
In the above equation, $U$ denotes the complex structure of
the torus with unprimed coordinates, whereas $S$ denotes
the complex structure of the torus with primed
coordinates.
We have so far always switched freely between cycles and forms using the 
natural duality. We now make this identification explicit at the
orientifold point:\footnote{
We have normalized the orientifold such that $\int_{T^2/Z_2}dx\wedge 
dy=1/2$ and $\int_{T^2}\wedge dx'\wedge dy'=1$.
}
\begin{center}
\begin{tabular}[h]{cc}
$e_1=$& $-2dy\wedge dy'$ \\ 
$e_2=$& $2dy\wedge dx'$ \\ 
$\alpha=$& $2dx\wedge dx'$ \\ 
$\beta=$& $2dx\wedge dy'$ \ .
\end{tabular}
\end{center}
Thus we have shown that the parameters $U$ and $S$ in
\begin{align}
\Omega_{SO(8)^4}=\frac{1}{2}\left(\alpha+Ue_2+S\beta-USe_1\right) \ ,
\end{align}
do indeed describe the complex structure of the base
and the fibre torus. 

The findings of this section represent one 
consistent identification of the torus cycles and
the $Z$-cycles that connect the four
blocks. It is possible to add appropriate
linear combinations of $e_1$, $e_2$, $\alpha$
and $\beta$ without destroying the mutual 
intersections and the intersection pattern
with the $16$ cycles in the four $SO(8)$ blocks.
What singles out our choice is the form of $\Omega_{SO(8)^4}$ 
in~(\ref{omegaso8}) as well as the $SO(16)$ that was implicitly defined 
in~(\ref{omegaso16}).

\section{D-Brane positions from periods and the 
weak coupling limit revisited}\label{dbp}

In this section, we study deformations away from the $SO(8)^4$ point.
To achieve this, we have to rotate the complex structure
such that not all of the vectors spanning the 
$D_4^{\oplus 4}$ lattice are orthogonal
to it. In other words, we want to add terms
proportional to the forms in (\ref{AtoD}) 
to $\Omega_{SO(8)^4}$.
To do this, we switch to an orthogonal basis defined by
\begin{align}
\tilde{E}_1&=E_1+e_2 ,\hspace{1cm} \tilde{E}_{I}=E_I,& I=2..4, 10..12 
\nonumber\\
\tilde{E}_9&=E_9+e_2 ,\hspace{1cm} \tilde{E}_{J}=E_J+e_1/2,& J=5..8, 13..16 \ .
\end{align}
As we will see, each $\tilde{E}_I$ is responsible for moving
only one of the D-branes when we rotate the complex structure to 
\begin{equation}
\Omega=\frac{1}{2}\left(\alpha+Ue_2+S\beta-\left(US-z^2\right)
e_1+2\tilde{E}_{I}z_I\right) \ .
\label{Omegagen}
\end{equation}
Here $z^2$ denotes $z_Iz_I$.
Note that all of the $\tilde{E}_I$ are orthogonal to
$\Omega_{SO(8)^4}$, so that we only have to change the coefficient
of $e_1$ to maintain the constraint $\Omega\cdot\Omega=0$. 

We can use the information about the length
of the blown-up cycles to compute the new positions of the 
branes.
Let e.g. $z_1\neq 0$: This gives the first cycle of block 
$C$, $C_1=-e_2-E_1+E_2$, the length $\Omega\cdot C_1=z_1$, so that we 
move one brane away from the O-plane. As a result, the $SO(8)$ at block 
$C$ is broken down to $SO(6)$. At the same time, the sizes of $Z_{AC}$ and 
$Z_{CD}$ are changed to 
\begin{align}
Z_{AC}\cdot\Omega=-\frac{1}{2}+z_1,\hspace{1cm}
Z_{CD}\cdot\Omega=-\frac{U}{2}-z_1 \ .\label{genha}
\end{align}
Thus we can move the brane from block $C$ onto block $A$ by
choosing $z_1=\frac{1}{2}$, or onto block $D$ by choosing 
$z_1=-U/2$. This can also be seen from the overall gauge group which is
$SO(6)\times SO(10)\times SO(8)^2$ for these two values of $z_1$. 

As we have seen in the last paragraph, $z_1$ controls the 
position of one of the four D-branes located at block $C$,
as compared to the position of the O-plane at block $C$.
If we let all of the $z_I$ be non-zero, we find the 
following values for the lengths of the cycles in block~$C$:
\begin{center}
\begin{tabular}[h]{c|c}
$C_I$ & $C_I\cdot\Omega$\\ 
\hline
$C_1$& $z_1-z_2$ \\ 
$C_2$& $z_2-z_3$ \\ 
$C_3$& $z_3-z_4$ \\ 
$C_4$& $z_3+z_4$
\end{tabular}
\vspace{-1.6cm}
\begin{equation}
\label{lenghtscI}
\end{equation}
\end{center}
\vspace{1cm}

\begin{figure}[!ht]
\begin{center}
\includegraphics[height=7cm,]{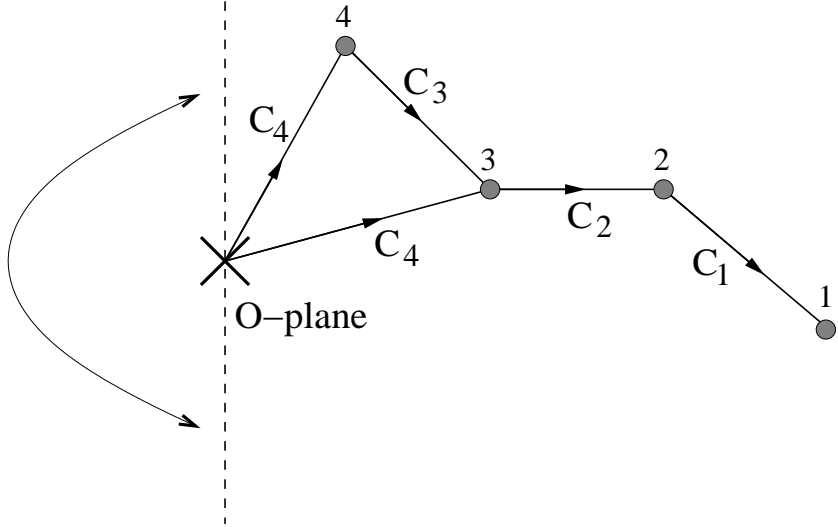}\caption{\textit{The
positions of D-branes on $\mathbb{C}/Z_2$ are measured 
by complex line integrals along the cycles $C_I$. 
As indicated in the picture, $\mathbb{C}/Z_2$ is obtained 
from the complex plane by gluing 
the upper part of the dashed line to its lower part.
Due to the presence of the O-plane, the line integral along 
$C_4$ has to be evaluated as
indicated by the arrows. Using (\ref{lenghtscI}), one can 
see that the positions of the branes are given by the $z_I$.}}
\label{dpos}
\end{center}
\end{figure}

To determine the D-brane positions, it
is important to note that the D-branes
are moving on a pillow, $T^2/Z_2$.
We thus use a local coordinate system equivalent
to $\mathbb{C}/Z_2$. It is centered
at the position of the O-plane of block $C$ 
at one of the corners of the pillow.
The intersections of the cycles with $\Omega$
are line integrals along the base part of the 
cycles $C_I$ (see Fig.~\ref{dpos}), multiplied 
by the line integral of their fibre part (which can
be set to unity locally). 
This is of importance for the length of $C_4$:
due to the orientation flip in the fibre when 
surrounding the O-plane, we have to evaluate both parts
of the line integral going from the O-plane to the
D-branes to account for the extra minus sign. 
This is indicated by the arrows that
are attached to the cycles in Fig.~\ref{dpos}. 
It is then easy to see that associating the $z_I$
with the positions of the D-branes yields the correct 
results. Note that one achieves the same gauge enhancement
for $z_3=z_4$ and $z_3=-z_4$, because for both values
one of the $C_I$ is collapsed, cf.~(\ref{lenghtscI}). 
Thus the D-branes labelled $3$ and $4$ have to be at the
same position in both cases, which fits with the
fact that $z_I=-z_I$ holds due to the $Z_2$ action.

By the same reasoning, the remaining $z_I$ give the 
positions of the other D-branes measured 
relative to the respective O-plane. 
For example, the moduli $z_5$ to $z_8$ give the positions of the 
D-branes of block $A$ (see (\ref{AtoD})).
We have also shown that we can connect the four blocks
by following the gauge enhancement that arises when
we move a brane from one block to another, cf.~(\ref{genha}). 
This means 
that we can also easily connect the four coordinate systems
that are present at the position of each O-plane. 
We have now achieved our goal of explicitly mapping 
the holomorphic 2-form $\Omega$ to the positions of the 
D-branes. For this we have used forms 
dual to integral cycles. These are the cycles that 
support the M-theory flux which can be used to stabilize the 
D-branes. By using our results, it is possible to derive the
positions of the D-branes from a given complex structure 
(unless the solution is driven away from the weak coupling 
limit). We thus view this work as an important step towards 
the explicit positioning of D-branes by M-theory flux. 

The geometric constructions of this article
only make sense in the weak coupling limit, 
in which the monodromies of 
the branes of F-theory are restricted to those 
of D7-branes and O7-planes. 
It is crucial that the positions of the D-branes
and the shape of the base torus factorize in the weak
coupling limit, $S\rightarrow i\infty$. The shape 
of the base torus is measured by 
multiplying the cycles $\alpha$, $\beta$, $e_1$ and $e_2$
with $\Omega$. The result is independent of the positions
of the branes in the weak coupling limit, as the only potential
source of interference is the $z^2$ in $\alpha\cdot\Omega=US-z^2$,
which is negligible as compared to $US$.
Thus the branes can really be treated as moving
on $T_2/Z_2$ without backreaction in the weak
coupling limit.

Certain gauge groups, although present in F-theory, do not 
show up in perturbative type IIB orientifolds and thus cannot 
be seen in the weak coupling limit. The lattice of forms 
orthogonal to $\Omega$ only has the structure of gauge groups 
known from type IIB orientifold 
models when we let $S\rightarrow \iu \infty$ in (\ref{omegagen}). 
This comes about as follows:
Starting from the $SO(8)^4$ point, we can cancel
all terms proportional to $W_I E_I$ in (\ref{omegagen})
when we are at finite coupling. In the limit
$S\rightarrow i\infty$, the fact that $\beta$ has
$S$ as its prefactor prevents the cancellation
of the term $W_I^2 E_I$ in $\Omega$. The 
presence of this term ensures that only
perturbatively known gauge enhancements 
arise.

\section{Summary and Outlook}

Our main points in this article were the implications of the weak coupling 
limit (which is necessary if one wants to talk about D7-branes moving in 
a background Calabi-Yau orientifold) and the description of D7-brane motion 
on $\mathbb{CP}^1$ in terms of integral M-theory cycles. 

Regarding the weak coupling limit, we have rederived the fact
that the deficit angle of a D7-brane is zero in a domain
whose size is controlled by the coupling. Furthermore, we
have pointed out `physics obstructions' that arise in the
weak coupling limit: The polynomial that describes the D7-brane
has to take a special form in the weak coupling limit, so
that some degrees of freedom are obstructed. We
have counted the relevant degrees of freedom explicitly for
F-theory on elliptically fibred spaces with base
$\mathbb{CP}^1\times\mathbb{CP}^1$ (which is dual to the 
Bianchi-Sagnotti-Gimon-Polchinski model) and with base $\mathbb{CP}^2$. 
From a local perspective, the obstructions arise at intersections
between D7-branes and O7-planes. They demand that D7-branes always 
intersect O7-planes in double-intersection-points (in the fundamental 
domain of the orientifold model; not in the double-cover Calabi-Yau, where 
this would be a trivial statement). The obstructions imply that a D7-brane
intersecting an O7-plane has fewer degrees of freedom than a corresponding 
holomorphic submanifold. It would be very interesting to analyse the 
implications of this effect explicitly in higher-dimensional models. 

The rest of this paper was devoted to an explicit 
discussion of the degrees of freedom of
F-theory on $K3$, corresponding to type IIB string
theory compactified on an orientifold of $T^2$.
We were able to construct and visualize the cycles that
control the motion of D7-branes on the base. 
Using these cycles, we have achieved a mapping between the
positions of the D-branes and the values of the
complex structure moduli. This was done using the singularities that 
arise at special points in the moduli space. 

We consider our analysis an important preliminary step for
the study of more realistic D7-brane models. First,
realistic models should contain fluxes to
stabilize the geometry and the D7-brane positions.
Using a higher-dimensional generalization of our map between cycles 
and D7-brane positions, it should be possible to determine explicitly the 
flux stabilizing a desired D7-brane configuration. A first 
step in this direction, from which one should gain valuable intuition, might 
be the reconsideration of F-theory on $K3\times K3$ with our tools.
Second, we want to consider F-theory compactifications on elliptically 
fibred Calabi-Yau manifolds of higher dimension. We expect new kinds of 
cycles and new configurations, for example cycles describing the 
recombination of D7-branes and configurations leading to gauge groups that 
are not simply laced. 

When F-theory is compactified on a six-dimensional 
manifold, the latter can be analysed with the methods used in this paper. 
The recombination of two intersecting D7-branes can be locally described 
as a deformation of the conifold, so that an $S^3$ emerges. This $S^3$ 
can be seen as the combination of a disc spanned in the tunnel connecting 
the two D7-branes and a horizontal cycle in the fibre torus. 

Although there exists a rich literature on elliptically fibred 
Calabi-Yau manifolds (see e.g.~\cite{Batyrev:1994hm, Bershadsky:1996nh, 
Klemm:1996ts, Candelas:1996su, Perevalov:1997vw,Candelas:1997eh,
Aspinwall:2000kf}), the direct visualization of the geometric objects 
achieved in this paper becomes harder in higher dimensions. We hope 
that the use of toric geometry methods will allow for a geometric 
understanding and simple combinatorial analysis in such cases.

\section*{Acknowledgments}
We would like to thank Tae-Won Ha, Christoph L\"udeling, Dieter L\"ust, 
Michele Trapletti, and Roberto Valandro for useful comments and discussions.

\end{document}